\documentclass{IEEEcsmag}

\usepackage[colorlinks,urlcolor=blue,linkcolor=blue,citecolor=blue]{hyperref}

\usepackage{wrapfig}
\usepackage{tabularray}
\usepackage{mathptmx} 
\usepackage{multirow}
\usepackage{graphicx}
\usepackage{rotating}
\usepackage{tcolorbox}
\usepackage{multicol}
\usepackage{float}
\usepackage{ulem}
\usepackage{cite} 
\usepackage{subcaption}
\usepackage{makecell}
\usepackage[table]{xcolor}

\usepackage{booktabs}

\usepackage{array}
\usepackage{amssymb}

\usepackage[table]{xcolor}
\usepackage{array}
\usepackage{tabularx}
\definecolor{markBlue}{HTML}{638FFF}
\definecolor{markOrange}{HTML}{FFB000}

\newcolumntype{Y}{>{\centering\arraybackslash}X}
\newcommand{\cmark}{\(\checkmark\)}
\newcolumntype{L}[1]{>{\raggedright\arraybackslash}p{#1}}
\newcolumntype{C}[1]{>{\centering\arraybackslash}p{#1}}
\usepackage[dvipsnames]{xcolor} 
\usepackage{placeins}
\usepackage{orcidlink}

\jvol{XX}
\jnum{XX}
\paper{8}
\jmonth{May/June}
\jname{IEEE Computer Graphics \& Applications}
\pubyear{2026}

\begin{document}

\sptitle{FEATURE ARTICLE}

\title{Vibe Analysis: Exploring LLM Adoption by Data Visualization Practitioners}

\author{{Shani C. Spivak} \orcidlink{0009-0001-7519-6584}}
\affil{Northeastern University, Boston, MA 02125, USA}

\author{{Aditi Krishna}
\orcidlink{0009-0004-1938-1525}}
\affil{Northeastern University, Boston, MA 02125, USA}

\author{{Mahsan Nourani}
\orcidlink{0000-0002-8823-9635}}
\affil{Northeastern University, Portland, ME 04101, USA}

\author{{Melanie Tory}
\orcidlink{0000-0002-6806-9253}}
\affil{Northeastern University, Oakland, CA 94613, USA}

\markboth{FEATURE ARTICLE}{FEATURE ARTICLE}

\begin{abstract}
\looseness-1Large language models (LLMs) are enticing in their promise to support data visualization (Vis) through faster and simpler workflows for data prep, analysis, and visualization creation. Yet LLMs are notoriously error-prone and not built for data visualization tasks. Few studies have explored LLM adoption among Vis practitioners. To fill this gap, we conducted semi-structured interviews with members of the Data Visualization Society, a global community of data visualization designers. Our findings show that Vis designers actively use LLMs for both creative and technical aspects of the visualization process. A new visualization workflow is emerging, a process we call vibe analysis, analogous to vibe coding. Some key challenges raised by participants parallel those of vibe coding, while others are Vis-specific, like gaps in Vis knowledge and chart verification. This work opens up opportunities for research combining LLM-mediated work with Vis tools that incorporate data visualization guidance, constraints, and best practices. 
\end{abstract}

\maketitle

\chapteri{C}onversational AI agents powered by Large Language Models (LLMs)\footnote{For brevity, we use the term \textit{LLM} in lieu of \textit{conversational AI agent} throughout this paper.} have the potential to enable a major leap forward in data visualization workflows. 
Visualization work can be complex; practitioners must balance various design goals with limitations related to data quality, access, computation, and tool knowledge.
Moreover, professionals across many roles are increasingly performing data analytics tasks even when data work is not their core responsibility. 
LLMs can support this work by offering design guidance or partially automating complex data pipelines. 
We present an interview study that explores whether LLMs are living up to this promise.

Although LLMs have potential to bridge the gap in skill and speed demanded of data visualization (Vis) tools and workers, they come with challenges and limitations, including hallucinations, deceptive behavior, and errors~\cite{yao2024llm}. Additionally, LLM use has been shown to impact how people approach, process, and remember task-related information~\cite{qian2024evolution, shaw2026thinking}. 
In Vis work, these impacts also relate to audience perception (how people interpret visualizations, data or topic biases, design issues, or viewer assumptions).  Data visualizations are often perceived as objective and factual, even when they are subjective, misleading, or incorrect~\cite{kosminsky2019belief}. 
Given the challenges of effectively detecting LLM errors, overreliance on LLMs in Vis work can be risky; undetected errors can lead users to costly misunderstandings and misinformed decisions.

So how \textit{are} data visualization practitioners using LLMs? 
Understanding current practices and challenges around the use of LLMs for data work is a critical first step to building better interaction tools, guidance, and governance.

To understand how visualization practitioners are using LLMs, we first conducted a survey advertised via the Data Visualization Society (DVS), a cross-functional, tool-agnostic community for sharing Vis design ideas and practices. We followed this with semi-structured interviews with a subset of participants to collect examples of LLM use, verification practices, and insights into participants' trust in and reliance on these tools.




We find that visualization practitioners already use LLMs across much of their work, in creative and surprising ways. 
Based on an analysis and synthesis of our interviews, we introduce the process of \textbf{vibe analysis}, which we define as an iterative analytic approach where users analyze data and generate visualizations by interacting with LLMs through natural language (NL) prompts rather than direct manual creation.
Vibe analysis is analogous to \textbf{vibe coding}~\cite{ge2025survey}, 
a programming paradigm where developers primarily generate code via interaction with code-generating LLMs through NL prompts, but with several data visualization-specific differences discussed in our findings. 
Our findings open up opportunities for future work on LLMs to better support Vis work through research into Vis-specific guidance for LLMs, Vis-specific LLM training for users, better integrations of LLM and no-code or low-code Vis tools, and adoption of a combination of LLM and Vis tools in ways that trade off the strengths and weaknesses of each. Our work contributes: 1) An introduction and in-depth description of vibe analysis, a conceptual model of a process by which visualization practitioners engage with LLMs for data work; 2) Information about how Vis practitioners are using LLMs, challenges and benefits in this process, and opportunities to facilitate this work; and 3) A call to action to the visualization community to think critically about where and how we apply LLMs and how we learn or train others to use them to present data.

\section{Related Work}

\subsection{User Preferences and Adoption of LLMs} Beyond user perceptions for data visualizations~\cite{kosminsky2019belief}, the impacts of LLM use on people who use them~\cite{shaw2026thinking}, and what data workers need from their tools~\cite{crisan2021passing,tory2019what}, previous work has also centered on user expectations and preferences for interactions with analytic chatbots.
For instance, Qian et al.~\cite{qian2024evolution} studied LLM adoption among practitioners at a large technology company, characterizing participants' perceptions, integration strategies, and reported usage scenarios. They showed how LLMs allow high-level insights from the start, reversing a previous bottom-up approach (i.e., data prep first, then insights). They also found that LLMs enabled automated insight discovery and data generation, though they cautioned that some of these changes could result in a shallower understanding. 

Closest to our work is an interview study by Schetinger et al.~\cite{schetinger2023doom} with 21 experts (across visualization, machine learning, art, and art history) to examine implications of generative models for visualization, identifying opportunities like rapid prototyping, while highlighting concerns about bias amplification and misleading visualizations. 
In contrast to their work, our study focuses specifically on visualization practitioners; moreover, our findings reveal additional insights about user trust, verification practices, and challenges in using LLMs for Vis. 

Although these studies contribute to a general understanding of data practitioner needs, none identify the approach we describe as vibe analysis.
Furthermore, we survey practitioners across countries, sectors, and roles and consider all Vis work, allowing for a more comprehensive view of adoption trends, user practices, and roadblocks. 

\subsection{Vibe Coding}
The way our participants described LLM use closely matches a collaborative, iterative process for software development known as \textit{vibe coding}.
In February of 2025, Andrej Karpathy introduced the term \textit{vibe coding} in a tweet, describing it as a way to abstract away from the programming language itself to develop software essentially through conversation with an LLM. Hailed for enabling the offloading of much of the code-writing behind software development, making development more accessible to non-coders, vibe coding is often described as opening up new creative possibilities to people who would not otherwise have the opportunity to code~\cite{ge2025survey}. Chou et al.~\cite{chou2025building} explored how software developers actually engage in vibe coding, revealing a variety of behaviors and degrees of reliance. They highlight how practitioners review, refine and correct, always contending with the stochastic nature of generation, what they describe as ``rolling the dice''. 

It can be expected, then, that other creative practices that depend on coding may benefit from a similar approach, but to our knowledge, no such process definition exists for data analysis or visualization. Visualization activities differ from coding in that Vis work also involves domain-specific tools, rules, heuristics, and constraints, many of which are learned by practitioners but are not necessarily part of model training or common knowledge to lay users of LLMs. Still, just as LLMs have seeped into the software development chain~\cite{nolan2026top,bindley2026tech}, they are making their impact in Vis.

While vibe coding has been explored across industry and academia~\cite{ge2025survey,chou2025building,pimenova2025good,geng2025exploring,sarkar2025vibe,umama2025llm}, to our knowledge, our work is the first interview study to provide a sector and role-agnostic view of how LLMs are being used in Vis processes (not just one part of the Vis pipeline or for a certain type of Vis). We offer a view of LLM adoption, including practices and limitations that can be used to improve interaction. Our work introduces \textit{vibe analysis} as a practice and describes its defining characteristics, benefits and challenges.

\section{Methodology}

Our goal was to characterize how LLMs are used by practitioners in their Vis work. Specifically, we wanted to understand current practices and challenges in the use of LLMs for visualization (RQ1) and in how practitioners verify LLM responses in this context (RQ2).

To address these questions, we designed a semi-structured interview for Vis professionals. To find these practitioners, we first developed a screening survey and distributed it via the Data Visualization Society (DVS). Survey participants could opt-in to be contacted for a paid interview. The study was approved by our Institutional Review Board and informed consent was obtained. Survey questions, interview protocol, and other study materials are available in the supplemental material.

\subsection{Participants \& Interview Protocol}
We manually reviewed responses to the screener survey to validate participants and ensure representation across diverse data work needs and tasks. 
This led to recruiting and interviewing 18 Vis practitioners (4 Female, 13 Male, 1 Prefer not to say) across 8 countries. Interviews were conducted between September and October, 2025.
See supplemental material for more information about interview participants.

\textbf{Interview Design and Protocol}
Interview questions built on the questions asked in the screening survey. 
We aimed to gather in-depth information about participants' use of LLMs in their Vis work through examples, questions about challenges, strategies, verification processes and concerns. We divided questions into three groups: 1) Questions about Vis Expertise (e.g., \textit{As related to data visualization, which of the following describe your role in the past year? Which tools do you tend to use for data visualization?}), 2) Questions about LLM use to contextualize their comfort level and general use (e.g., \textit{For what part of the data visualization process do you use Large Language Models (LLMs)?}), and 3) Demographic Questions (e.g., age range, gender). We included questions about trust and verification because we were interested in practitioners’ perspective on these, but had no guiding framework for this and all qualitative coding was bottom-up. Please see Supplemental Material for the full survey and interview guide.


Two researchers on our team carried out the interviews virtually over Microsoft Teams.  Average interview duration was 28m (SD: 8m). Sessions were recorded and transcribed. We anonymized all transcriptions before analysis. Researchers also took notes of the most important points in the discussion, which were used in one case where the recording failed.
Participants who completed the interview received a $\$30$ (USD) gift card.

\subsection{Qualitative Analysis}
For each transcript, we applied inductive thematic coding ~\cite{naeem2023step}. First, two researchers independently conducted a round of open coding, followed by a session to reconcile and consolidate themes. The coding process included a mix of in vivo codes and researcher-denoted concepts. New codes were developed throughout transcript review. Finally, both researchers independently performed closed coding and then results were merged. There were some instances where only one coder noted a finding, but otherwise any differences were resolved through discussion. 
Several themes were identified as part of focused coding and additional review, which we discuss next in the Findings Section. 

Crisan et al.~\cite{crisan2021passing} inspired us to examine where in the Vis pipeline participants were using LLMs. However, for Vis pipeline components, we adopted naming conventions and groupings of higher and lower order processes based on descriptions provided by our participants; this ensures that the pipeline components presented here directly reflect the results of our interviews.

\section{Findings}
\label{findings}

\begin{figure*}
    \centering
    \includegraphics[width=1\linewidth]{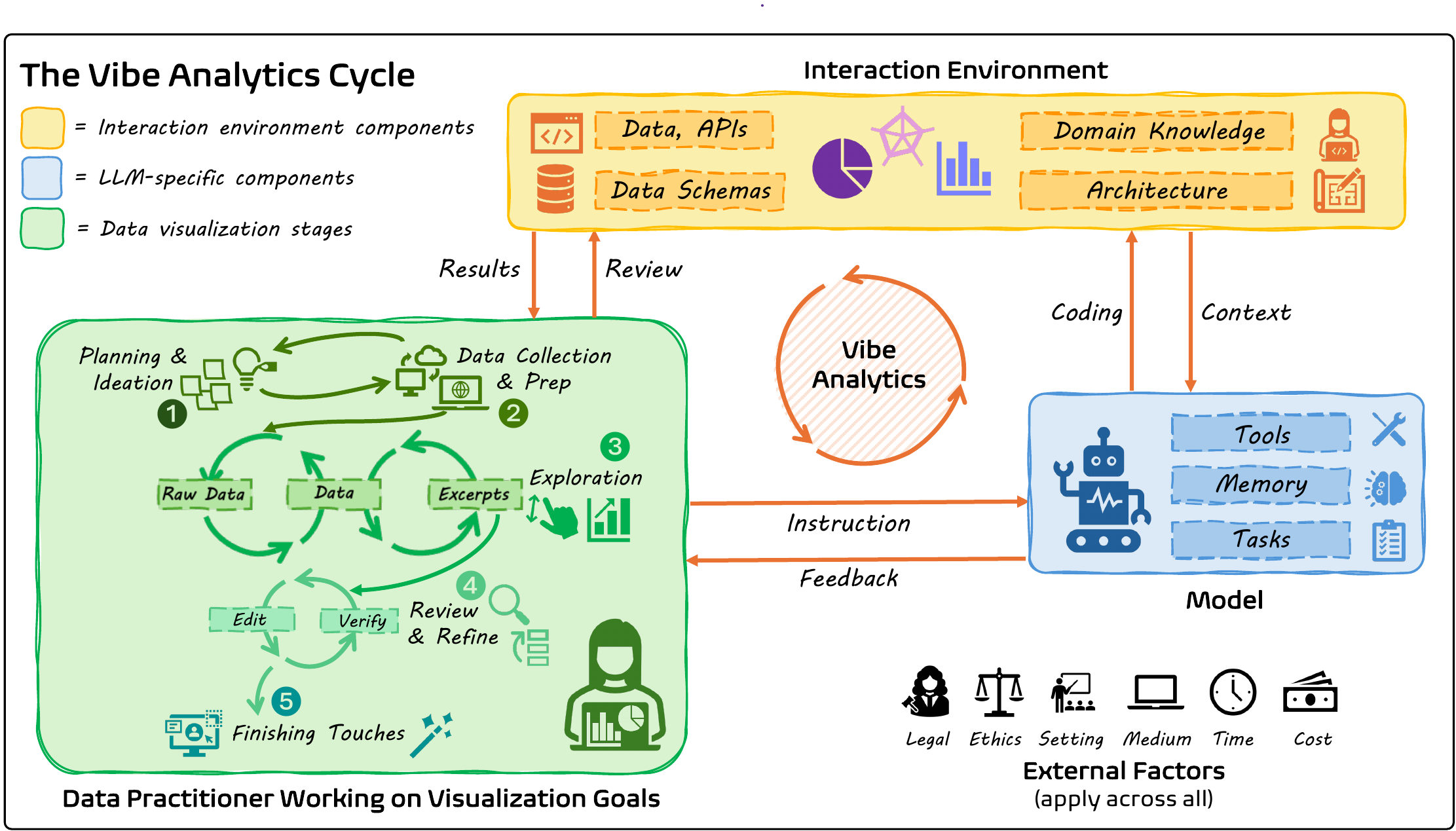}
    \caption{Conceptual model of vibe analysis with portions inspired by the data storytelling lifecycle~\cite{riche2018analysis} and the vibe coding cycle~\cite{ge2025survey}. This is an author-developed synthesis informed by the interview analysis and these existing models.}
    \label{fig:vibe}
\end{figure*}

Our participants described an iterative process of working with LLMs to accomplish data tasks across various stages of the visualization pipeline. Below, we characterize what we call \textbf{vibe analysis} (see \autoref{fig:vibe} for a conceptual model). In the subsections below, we first describe vibe analysis and then discuss themes around the changing nature of visualization work, strategies employed by participants, and challenges they brought up. The scope of Vis work described here reflects what participants described in interviews and framing from related research~\cite{crisan2021passing,riche2018analysis}, but this is inherently non-comprehensive, as other aspects of data analysis and visualization are also likely to be impacted by LLM use. 
Throughout the findings, ratios indicate the number of interviewees out of 18, unless otherwise noted. These values are descriptive indicators, but should not be taken as estimates of prevalence among visualization practitioners more broadly. Information about specific LLMs participants mentioned using is provided in~\autoref{fig:tools} and more detailed information about which tools participants use is in Table 3.1 of the supplemental material. 

\begin{figure}
    \centering
    \includegraphics[width=0.7\linewidth]{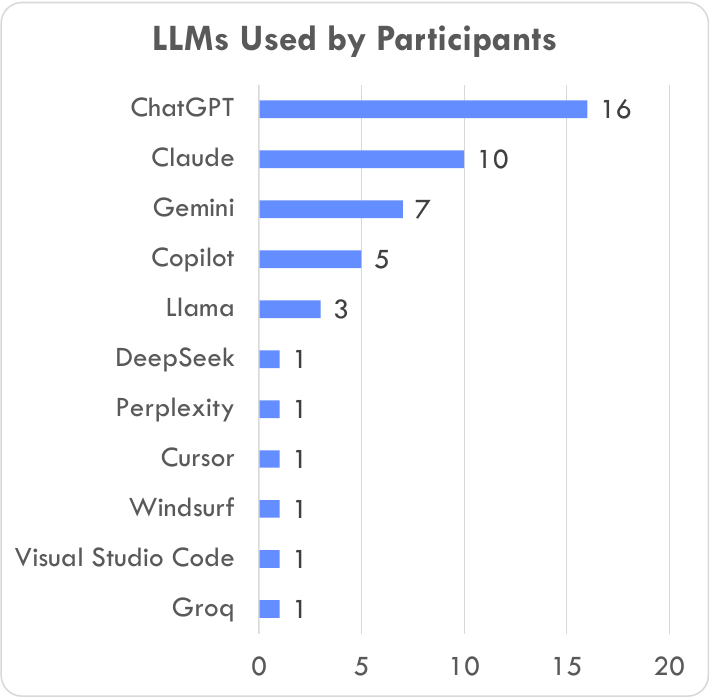}
    \caption{LLM tools our participants reported using}
    \label{fig:tools}
\end{figure}

\subsection{Vibe Analysis}
One of our participants described their use of LLMs as: \textbf{P9}: ``\textit{[...] comparable to me taking somebody's data, mocking it up really quick and sending them a screenshot. [...] 
    another use case is not vibe coding, but \textbf{vibe modeling} [when] I want to see what a fake dataset will look like.}''

\noindent Like this participant, many others described working with LLMs in loosely organized sessions, prompting LLMs with goals rather than specific directions.
Most participants (11/18) used anthropomorphic language in relation to LLMs, and indicated a kind of collaborative design process that differs from more traditional Vis tool use in which data is organized and visualized according to the tool-specific processes (e.g., Tableau's interface or Excel's chart options). Practitioners used vibe analysis to quickly get a workable overview or chart(s), often using LLMs to brainstorm (ideate), wrangle, explore, and visualize data. 

Vibe analysis is related to \textbf{vibe coding}, 
a coding approach that relies on LLMs, allowing programmers to generate working code by providing natural language descriptions rather than manually writing it~\cite{ge2025survey}. We use \textbf{vibe analysis} as an umbrella term for the full process of LLM-mediated visualization work.
Like vibe coding, vibe analysis involves an iterative process of completing data tasks and creating visualizations, where much of the work is done through NL prompts to an LLM rather than direct specification. Humans shift from direct code-writing / pipeline specification to articulating goals, curating content, and evaluating quality. 

Figure~\ref{fig:vibe}, adapted from Ge at al.~\cite{ge2025survey}, shows a conceptual model of vibe analysis that developed as a result of our coding process. As in any visual analytics process, the practitioner works through a series of analytics tasks (green box, bottom left) including planning \& ideation, data collection \& prep, exploration, review \& refinement, and finishing touches. Any of these activities might be supported by, or primarily conducted through, an LLM assistant. The iterative vibe analytics loop (center of figure) involves human specification of NL instructions to an LLM, code creation by the LLM, and review of results and/or code by the analyst. This may all happen within the LLM's UI, may be orchestrated via more than one LLM, or may occur in concert with traditional Vis tools (e.g., Tableau, Power BI, Excel) and separate code environments (e.g., Observable, Svelte). Various external factors (bottom right) contextualize this process and may influence how it unfolds. The central cycle in orange arrows includes the cyclical and iterative actions represented by the other orange arrows---the user providing instructions to the LLM and reviewing project components, the LLM providing feedback and questions to the user and producing code, and the project's interaction environment providing both context for the LLM and a space to review and iterate on results.

In comparison to the vibe coding conceptual model~\cite{ge2025survey}, our vibe analytics conceptual model includes the Vis pipeline, and updates project context to more closely mirror Vis-specific constraints like data access and format restrictions and architecture needs. We also incorporate external factors noted in  Riche et al.'s data storytelling process~\cite{riche2018analysis}
because all of these factors came up in interviews; this includes deadlines, LLM costs, data restrictions, and applicable law and policy. 
Components of the vibe analytics process, as observed in our interviews, are described in detail in the sections below. 

\subsubsection{Visualization Goals}
Practitioners noted use of LLMs throughout the Vis development process, illustrated by activities in the green box of \autoref{fig:vibe}. While the phases and analytics tasks align with existing approaches to Vis work, participants described a shift in the \textit{way} they work as a result of using LLMs.
In ~\autoref{tab:pipeline}, we break down aspects of the Vis pipeline grouped by stage (planning and ideation, data collection and prep, exploration, review and refine, and finishing touches), and detail these higher order processes with the percent of participants who mentioned using LLMs for the underlying tasks.

\definecolor{headPurple}{HTML}{E59FDD}
\definecolor{headPink}{HTML}{FFCFFD}

\definecolor{llmBlue}{HTML}{628FFF}
\definecolor{llmOrange}{HTML}{FFB000}
\definecolor{llmCheck}{HTML}{4E6F8D}

\definecolor{rowVeryLightBlue}{HTML}{EEF4FF}
\definecolor{rowPaleBlue}{HTML}{F6F9FF}
\definecolor{rowLightBlue}{HTML}{DEE7FF}
\definecolor{rowMediumBlue}{HTML}{98B4FF}
\definecolor{rowBlueThree}{HTML}{D4E1FF}
\definecolor{rowBlueFour}{HTML}{D9E1FA}
\definecolor{rowBlueFive}{HTML}{BCCEFF}
\definecolor{rowBlueSix}{HTML}{CDDBFF}

\definecolor{rowCream}{HTML}{FFF6E3}
\definecolor{rowCreamTwo}{HTML}{FEF1D3}
\definecolor{rowCreamThree}{HTML}{F8E8C1}
\definecolor{rowGold}{HTML}{FFD062}
\definecolor{rowGoldTwo}{HTML}{FFB50D}
\definecolor{rowGoldThree}{HTML}{FFE5AB}

\newcommand{\bluecell}{%
  \cellcolor{llmBlue}\textcolor{white}{\cmark}%
}

\newcommand{\orangecell}{%
  \cellcolor{llmOrange}\textcolor{llmCheck}{\cmark}%
}

\newcommand{\shadecell}[2]{%
  \cellcolor{#1}\textbf{#2}%
}

\newcommand{\phead}[1]{%
  \cellcolor{white}\textcolor{black}{\textbf{#1}}%
}

\newcommand{\countcell}[2]{%
  \cellcolor{#1}\textcolor{white}{\textbf{#2}}%
}

\begin{table*}[!t]
\centering
\caption{Parts of the Vis pipeline for which participants mentioned using LLMs. We distinguish between processes which are mostly \textcolor{blue}{technical (in blue)} and those which are mostly \textcolor{orange}{creative (in orange)}. Participants are ordered from the left with those using LLMs across more processes to those using them for fewer on the right.}
\label{tab:pipeline}

\setlength{\tabcolsep}{2pt}
\renewcommand{\arraystretch}{1.12}

\resizebox{\textwidth}{!}{%
\begin{tabular}{
    @{}
    L{2.2cm}|
    L{2.7cm}|
    C{2.1cm}|
    *{18}{c|}
    @{}
}
\toprule

\textbf{\shortstack[l]{Higher-order\\processes}}
&
\textbf{\shortstack[l]{Lower-order\\processes}}
&
\textbf{\shortstack{Count of\\participants \\ using LLMs}}
&
\phead{P4}
&
\phead{P7}
&
\phead{P14}
&
\phead{P6}
&
\phead{P10}
&
\phead{P11}
&
\phead{P17}
&
\phead{P5}
&
\phead{P3}
&
\phead{P9}
&
\phead{P15}
&
\phead{P12}
&
\phead{P16}
&
\phead{P18}
&
\phead{P1}
&
\phead{P2}
&
\phead{P8}
&
\phead{P13}
\\
\midrule

\multirow{2}{2.55cm}{Planning \& \\ Ideation}
&
\shadecell{white}{Planning}
&
\shadecell{rowVeryLightBlue}{2/18}
&
&
&
&
\bluecell
&
&
\bluecell
&
&
&
&
&
&
&
&
&
&
&
&
\\

&
\shadecell{white}{Ideation}
&
\shadecell{rowGold}{11/18}
&
\orangecell
&
&
\orangecell
&
\orangecell
&
\orangecell
&
\orangecell
&
\orangecell
&
\orangecell
&
\orangecell
&
&
\orangecell
&
\orangecell
&
&
\orangecell
&
&
&
&
\\
\midrule

\multirow{6}{2.55cm}{Data Collection \\ \& Prep}
&
\shadecell{white}{Definitions}
&
\shadecell{rowPaleBlue}{1/18}
&
\bluecell
&
&
&
&
&
&
&
&
&
&
&
&
&
&
&
&
&
\\

&
\shadecell{white}{Data Collection}
&
\shadecell{rowLightBlue}{4/18}
&
&
\bluecell
&
\bluecell
&
\bluecell
&
&
&
&
&
&
&
\bluecell
&
&
&
&
&
&
&
\\

&
\shadecell{white}{Data Generation}
&
\shadecell{rowPaleBlue}{1/18}
&
&
&
&
&
\bluecell
&
&
&
&
&
&
&
&
&
&
&
&
&
\\

&
\shadecell{white}{Data Labeling}
&
\shadecell{rowVeryLightBlue}{2/18}
&
&
&
&
&
&
&
&
&
&
\bluecell
&
&
&
&
&
&
&
&
\bluecell
\\

&
\shadecell{white}{Data Wrangling}
&
\shadecell{rowMediumBlue}{13/18}
&
\bluecell
&
\bluecell
&
&
\bluecell
&
\bluecell
&
&
&
\bluecell
&
&
\bluecell
&
\bluecell
&
\bluecell
&
\bluecell
&
&
\bluecell
&
\bluecell
&
\bluecell
&
\bluecell
\\

&
\shadecell{white}{Examples}
&
\shadecell{rowCream}{2/18}
&
\orangecell
&
&
\orangecell
&
&
&
&
&
&
&
&
&
&
&
&
&
&
&
\\
\midrule

\multirow{6}{2.55cm}{Exploration}
&
\shadecell{white}{Exploration}
&
\shadecell{rowMediumBlue}{12/18}
&
\bluecell
&
\bluecell
&
&
\bluecell
&
\bluecell
&
\bluecell
&
\bluecell
&
&
\bluecell
&
&
\bluecell
&
\bluecell
&
\bluecell
&
\bluecell
&
&
&
\bluecell
&
\\

&
\shadecell{white}{Experimentation}
&
\shadecell{rowCreamTwo}{3/18}
&
&
\orangecell
&
&
&
&
&
&
&
&
\orangecell
&
&
&
&
\orangecell
&
&
&
&
\\

&
\shadecell{white}{Prototyping}
&
\shadecell{rowCreamThree}{5/18}
&
&
&
\orangecell
&
\orangecell
&
\orangecell
&
&
&
&
&
\orangecell
&
\orangecell
&
&
&
&
&
&
&
\\

&
\shadecell{white}{Analysis/Modeling}
&
\shadecell{rowBlueThree}{5/18}
&
&
\bluecell
&
&
&
&
&
\bluecell
&
&
\bluecell
&
&
&
&
\bluecell
&
&
&
&
&
\bluecell
\\

&
\shadecell{white}{Visualization Design}
&
\shadecell{rowGoldTwo}{17/18}
&
\orangecell
&
\orangecell
&
&
\orangecell
&
\orangecell
&
\orangecell
&
\orangecell
&
\orangecell
&
\orangecell
&
\orangecell
&
\orangecell
&
\orangecell
&
\orangecell
&
\orangecell
&
\orangecell
&
\orangecell
&
\orangecell
&
\orangecell
\\

&
\shadecell{white}{Insights}
&
\shadecell{rowBlueFour}{3/18}
&
&
&
&
&
&
\bluecell
&
&
&
\bluecell
&
&
&
&
&
&
&
&
\bluecell
&
\\
\midrule

\multirow{5}{2.55cm}{Review \& Refine}
&
\shadecell{white}{Feedback}
&
\shadecell{rowLightBlue}{4/18}
&
&
&
\bluecell
&
&
&
&
\bluecell
&
\bluecell
&
\bluecell
&
&
&
&
&
&
&
&
&
\\

&
\shadecell{white}{Programming}
&
\shadecell{rowBlueFive}{8/18}
&
\bluecell
&
\bluecell
&
\bluecell
&
\bluecell
&
\bluecell
&
&
&
\bluecell
&
&
&
&
&
&
\bluecell
&
&
\bluecell
&
&
\\

&
\shadecell{white}{Debugging}
&
\shadecell{rowBlueSix}{6/18}
&
\bluecell
&
&
\bluecell
&
&
&
\bluecell
&
&
\bluecell
&
&
&
&
&
&
&
\bluecell
&
\bluecell
&
&
\\

&
\shadecell{white}{Reformatting}
&
\shadecell{rowVeryLightBlue}{2/18}
&
&
&
&
&
&
&
\bluecell
&
&
&
&
&
&
&
&
\bluecell
&
&
&
\\

&
\shadecell{white}{Verification}
&
\shadecell{rowLightBlue}{4/18}
&
&
\bluecell
&
&
&
&
\bluecell
&
&
\bluecell
&
&
&
&
&
\bluecell
&
&
&
&
&
\\
\midrule

\multirow{3}{2.55cm}{Finishing Touches}
&
\shadecell{white}{Polish}
&
\shadecell{rowCreamTwo}{3/18}
&
\orangecell
&
&
\orangecell
&
&
&
&
\orangecell
&
&
&
&
&
&
&
&
&
&
&
\\

&
\shadecell{white}{Language/Storytelling}
&
\shadecell{rowGoldThree}{6/18}
&
&
\orangecell
&
\orangecell
&
&
\orangecell
&
\orangecell
&
\orangecell
&
&
&
&
&
\orangecell
&
&
&
&
&
&
\\

&
\shadecell{white}{Website Support}
&
\shadecell{rowPaleBlue}{1/18}
&
&
&
&
&
&
&
&
&
&
\bluecell
&
&
&
&
&
&
&
&
\\
\midrule

\multicolumn{3}{l}{
    \textbf{Count of processes where LLMs are used}
}
&
{9}
&
{9}
&
{9}
&
{8}
&
{8}
&
{8}
&
{8}
&
{7}
&
{6}
&
{6}
&
{6}
&
{5}
&
{5}
&
{5}
&
{4}
&
{4}
&
{4}
&
{4}
\\

\bottomrule
\end{tabular}%
}
\end{table*}

We found that most interviewees used LLMs for visualization design (17/18),  data wrangling (13/18), exploration (12/18), and ideation (11/18).
Some participants reported using LLMs for analysis (5/18), insight discovery (5/18), programming (8/18), debugging or troubleshooting support (6/18), and language or storytelling support (6/18). These findings demonstrate that Vis practitioners use LLMs for both creative and functional aspects of the visualization pipeline, which was somewhat surprising to us; we anticipated that practitioners would prefer to offload technical and repetitive tasks to LLMs, basically vibe coding, while handling creative tasks themselves.

Interestingly, use of LLMs for 
finishing touches (like reformatting or polish) were less common than for earlier stages of visualization work. In addition, LLMs were also underutilized for some repetitive activities, such as planning, data labeling, data generation, and reformatting. We delve further into each stage below.
\\\\
\noindent\textbf{Planning \& Ideation:} At this stage, LLMs were used for preparatory tasks including organizing and brainstorming in the beginning of the pipeline. 
Overall, 11/18 of participants noted using LLMs in early phases, mostly for creative ideation but sometimes for exploring how to approach analysis and design. For example, \textbf{P4} described using LLMs during the creative process: \textit{``[...] 
I might ask for \textbf{suggestions on which chart types might suit a dataset} or how to make a visualization more engaging [...]''}. 
\textbf{P6} described prompting the LLM with starter information and asking for advice on an analysis plan: ``\textit{I write a small prompt [...] maybe 10 lines just stating what do I want and [...] I ask the chat to \textbf{improve this small prompt to make it into a plan} for some like visualization or code change.}''

Some participants did raise concerns about the potential for a LLM to warp or constrain creativity (\textbf{P9}: \textit{``\textbf{Never ideation}. I don't want it to corrupt my brain.''}, \textbf{P1}: \textit{``\textbf{I never use [it] for giving me ideas} about [...] charts because I know they are giving four basic charts.''}).
\\\\
\noindent\textbf{Data Collection \& Prep:} LLMs were used to guide or manage data tasks including exploring data definitions and examples, data collection or generation, data labeling, and data formatting. 14/18 of participants mentioned using LLMs for data collection and prep. Most participants described using LLMs for data wrangling tasks, such as data cleaning, processing, enrichment, and prep (\textbf{P1}: \textit{``Yes, very much a time saving thing [...]
\textbf{cleaning is one of the most important parts}''}). Some participants also noted using LLMs in data collection (4/18) and synthetic data generation (1/18). Other tasks included asking for definitions, examples, and data labels.
\\\\
\noindent\textbf{Exploration:} This includes using LLMs to conduct exploratory data analysis, statistical analysis and other data modeling, along with experimentation and prototyping. Other key LLM tasks at this stage include insight generation and visualization design. This stage had the highest rate of LLM use at 17/18. Many participants noted a preference for using LLMs for more complex visualizations (\textbf{P13}:\textit{ ``I have tried to use LLMs to create data visualizations. I thought it would be easy here for me for example to \textbf{make like a map} because you know a bar chart I can do just as quickly in myself;}'' \textbf{P7}:\textit{
``[...] \textbf{to build more complex data visualization} and I mean like dashboards or web-based interface when you need to navigate
[...] to build a product or a final output.''}). Also common was exploration to identify trends and better understand the data.  
Some participants noted offloading much of the analytic work to LLMs 
(\textbf{P3}: \textit{``\textbf{the analysis} is being taken care of by the LLM.''}), while others used it for rapid prototyping and other experimentation (\textbf{P10}:\textit{
``I create \textbf{rough outlines of things} using LLM-based sketches for moving towards production.''}). 
\\\\
\noindent\textbf{Review \& Refine:} In one of the last stages of the pipeline, participants noted use of LLMs for feedback, reformatting, programming and debugging support. This is also where those who use LLMs for verification do so. Overall, 12/18 of participants noted using LLMs in this stage, with support in programming (8/18) as the most frequently noted. For instance, \textbf{P5} 
reported using AI to get,
``\textit{[...] \textbf{code snippets}, maybe to \textbf{brush up my code or even get a much better code snippet} that helps me [...]}''. Others mentioned using LLMs to seek feedback on visualization design (4/18), for debugging or troubleshooting (6/18), and for checking its own or another model's work (4/18). 
\\\\
\noindent\textbf{Finishing Touches:} In these last steps, LLMs are used to improve language and storytelling, add polish to an otherwise finished product, or support website integration. Only 8/18 of participants noted using LLMs for finishing touches, with language and storytelling support as the most common (6/18). This usage reflects both language support for non-native English speakers (\textbf{P7}: \textit{``I also find it useful to kind of \textbf{write the text in English} and pass through an LLM to \textbf{improve fluency}''}) and for storytelling  
(\textbf{P17}: `\textit{`So I believe that with the use of [LLMs] I can improve my visual[s] and also \textbf{improve my storytelling}.''}). We speculate that lower usage of LLMs in applying finishing touches might be due to the lack of consistency and precision. \textbf{P13} described a challenge with getting the LLM to appropriately apply finishing touches: \textit{``I asked it to shade, you know, the following states, and it did it brilliantly, except it forgot one. And so then when I asked it to add that one, it added that one plus the Eastern seaboard. And then I'm like, well, wait, what happened? And then it like uncolored the Eastern seaboard and put in all the Canadian provinces.''} \textbf{P13} eventually had to give up, \textit{``And so despite many attempts to use the [LLM] to make this map, I ended up having to hand do it with dots [...].
So it just didn't look as nice, but it was at least accurate.''}
This example illustrates an important opportunity to improve LLM-supported vis tools to maintain a constrained and consistent behavior. It also highlights a concern about the stochastic nature of LLMs -- what~\cite{chou2025building} et al. describe as rolling the dice -- using prompt after prompt of similar intent to try to get to a good outcome knowing they'll get something new each time.

\subsubsection{The Interaction Environment}
The vibe cycle, model, and constraint elements in Figure~\ref{fig:vibe} bear many similarities to vibe coding. 
We distinguish between constraints (which are project-based specifications and project team domain knowledge related to the interaction environment and model) and external factors (which include policy, time and other factors affecting, but not specific to a project, like data privacy regulations and organizational licenses). 
Recognizing the role of data sources, tools, and schemas as first class citizens in Vis, we draw together a set of constraints that may apply to any given Vis project. 

\textbf{Constraint 1: Data, APIs.} Data sources and APIs impact project requirements and constrain analytic options. 
\textbf{P14:} ``\textit{The software that is created is using R and we connect it with the models through the API}'' \textbf{P15:} ``\textit{I've used it for pulling data from APIs}''
\textbf{Constraint 2: Data Schemas} define how data is related, stored, and organized, therefore constraining analytic options. 
\textbf{P14:}
 ``\textit{...send the model the structure of the database, not the database, just the schema}''
\textbf{Constraint 3: Domain Knowledge.} we adopt Ge et al.'s~\cite{ge2025survey} description of domain knowledge, as documentation, specifications, and best practices. 
\textbf{P18:} ``\textit{The transfer of information in hand to the machine gets difficult sometimes [...] 
For example, if you would ask a fellow lab mate [to] explain them the same thing, it's much easier because they would be much more aware of the context that's around it.}''
\textbf{Constraint 4: Architecture} where vibe coding lists execute environment, we instead opt for architecture to 
include the various applications (Excel, Tableau, Power BI), coding languages and environments that may be integrated. 
\textbf{P1:} ``\textit{Copilot because I have 365 subscription. So for Excel thing copilot I found is pretty much they can work with me for five hours, no questions.}''

\subsubsection{The Model}
We adopt Ge et al.'s~\cite{ge2025survey} descriptions of vibe coding components: 1) \textbf{Tools:} Definitions and signatures of callable tools (compilers, testing frameworks, version control);
2) \textbf{Memory:} Historical interaction memory (multi-turn dialogue context, prior decision records); and
3) \textbf{Tasks:} Current tasks (pending actions, task queue, execution status). 

\subsubsection{External Factors}
\label{sec:external}
Participants described many factors that influenced the vibe analytics process (see Figure~\ref{fig:vibe}, bottom right), mirroring the external factors identified by Riche et al.~\cite{riche2018analysis} for data storytelling.
As Riche et al. noted, target audience considerations impact each step of the process, but are distinct from analytic and creative processes. Our interviews suggest external factors also impact constraints and the model; participants sometimes noted legal and cost considerations to their use of specific models or versions.\\
\textbf{External Factor 1: Legal.} May include copyright or data use restrictions, among other Vis-specific legal implications like privacy protections. \textbf{P13:} ``\textit{We have an internal copilot that only has our data [...] So that's what I'm using. Both because of the privacy issues, but also just frankly, because we're restricted, I can't use anything else here.}''\\ 
\textbf{External Factor 2: Ethics.} Ethical considerations related to the data being shared or the potential for misuse or misinterpretation. Misleading charts are a concern for the Vis community~\cite{mcnutt2020surfacing}.
\textbf{P9:} ``\textit{I think the parts that worry me are when you start asking it things [...] about people where [...] the whole thing is built on a dataset of racist humans and so the model itself will be racist in ways that no one will catch and it'll just [...] keep kind of propagating that in ways that are harder to identify.}''\\
\textbf{External Factor 3: Setting.} What is the context of the work? Is it part of a presentation or interactive exhibit? Setting impacts design choices.  \textbf{P10:} ``\textit{You can find lots of examples on my blog where I have shared both snapshots and interactive versions of what I've created with LLMs.}'' \textbf{P15:} ``\textit{I'm mostly a static person. I'm mostly building out [of] Illustrator or Excel, and it's allowed me to expand on that into interactive pieces, which I just haven't been able to build.}''\\
\textbf{External Factor 4: Time.} Will the audience be guided synchronously or will they view a project asynchronously? This may impact how features are presented and annotated. \textbf{P18:} ``\textit{Let's say if there's a lot of variables included in [...] multimodal data that we have collected, then we would love to have help from LLM there. What is the best way to represent it so that people can get the [...] simplicity and the also the interaction between different modalities that we are trying to use?}''\\
\textbf{External Factor 5: Cost.} May include costs to access data, licensing, or proprietary Vis tools.  
\textbf{P18:} ``\textit{So I think most of the newer models [...] have a similarity in terms of intelligence [...] especially in the free tier. We don't have a paid version of any of these LLMs, so [...] if the one free tier expires then [...] we move to the other LLM.}''\\ 
\textbf{External Factor 6: Medium.} Choice of medium (e.g. scrollytelling site\footnote{Refers to a website where visualizations, text, and media updates as the user scrolls.}, dashboard, static chart) has impacts across the Vis process. \textbf{P4:} ``\textit{I found Claude useful when I need more narrative supports like writing [...] descriptions [...] 
because it's better at longer structured text and I use Gemini for design related brainstorming.}''


\subsection{Vibe Analysis and the Changing Nature of Data Work}

Our interviews indicate that LLMs are changing the way in which visualization and analytics tasks are accomplished by data practitioners.
Overall, our interviewees tended to use vis tools when they had specific details to visualize or needed simple charts they could build quickly themselves, and used LLMs for more complex, vague, or text-heavy tasks. 
5/18 of participants explicitly noted a \textbf{shift in the kind of work} they did or how they did it. Examples include shifting from using vis tools to coding (\textbf{P2}: \textit{``I would say it helped me \textbf{improve my skills} and save my time and it changed [...] some working habits like \textbf{shifted from a freely developed software to coding languages}.''}) or from coding to prompting (\textbf{P6}: \textit{``I can do more in less time, but the task has \textbf{moved away from coding skills to more prompting}, to creation, to select[ing] what is useful from the elements.''}). Changes in how work is done with AI mean that practitioners are abstracting away many fine-grained details of design and development, representing 
a shift in how visualizations are crafted. 

Participants described many benefits to using LLMs in data work, including increased efficiency (14/18), insight discovery (11/18), and expanded capabilities (9/18). Efficiency was often described in terms of taking over or helping to speed up parts of a process. For example, \textbf{P6} used LLMs to bootstrap code writing, \textit{``\textbf{I don't do anything from scratch} any longer,''} and \textbf{P11} used LLMs for brainstorming, \textit{``Previously this ideation process \textbf{could take months} 
[...] \textbf{Now the same thing can be done in 15 days} or one month.''}

Supporting insight discovery was described as complementary to human practices, often generating ideas that users may not have identified independently (\textbf{P13}: \textit{``There was one framing that [the LLM] did where I was like, 
\textbf{I don't know if I would have come up with [that] on my own}.''}). Expanded capabilities allowed users to code in unfamiliar languages, use tools without training, or develop products they could not otherwise, such as P15 being able to build interactive tools that they were unable to create before. 
Similar skill expansion has been observed in the use of vibe coding, enabling non-developers to build working applications~\cite{ge2025survey}. This is largely beneficial, but 
could lead to technical debt (see the Challenges Section).

Participants' perception of the impact of LLMs on their jobs and the nature of work echo sentiments shared by developer communities and tech sector leadership~\cite{nolan2026top,bindley2026tech}.
The \textbf{fear of being replaced} was frequently noted (\textbf{P5}: \textit{``So yeah, I actually worry that \textbf{one day this language model would be good enough to take over my job}''}).
Additionally, 5/18 of participants felt \textbf{compelled to use AI} for work, either because their employers told them as much, or because they would fall behind otherwise (\textbf{P4}: \textit{``But I think \textbf{people who don't learn to use them, they would be at a really great disadvantage}.''}
\textbf{P5}: \textit{``\textbf{It doesn't seem practical} in this day and time to try \textbf{to get information and visualize it without the use of the language model}''}). This corroborates recent reporting that firms are pushing their employees to use AI in their work~\cite{bindley2026tech}.

\subsection{Vis Practitioners' Strategies}
\label{sec:strategies}

\textbf{Changing Prompting and Interactions with LLMs} (Pipeline Phase: all). Participants commonly noted error-handing strategies through a change in their interactions with the model. Prominent themes included giving the LLM more context or guidance (9/18), starting over (8/18), breaking tasks down into steps or chunks (6/18), and refining the prompt (6/18). About 4/18 ask the LLM to check its own work and 3/18 mentioned they ask the LLM for explanations. 
\\\\
\noindent\textbf{Error Prevention} (Pipeline Phase: all). Practitioners reported strategizing to prevent potential LLM errors, for example by having a plan before getting started (4/18), using a Git repository for version control to enable easy, stepwise corrections (2/18)\footnote{We did not ask about Github, it came up organically in only 2 of the 18 interviews.}, being very specific in prompting (5/18), describing the data (4/18) or providing a sample (1/18), and asking for pseudocode or code instead of trying to get directly to a visualization (1/18). Others opt to only use LLMs for repetitive tasks (e.g. providing the LLM with their own code and guidance on how to repeat a process) (3/18) or just for rough sketches or starting points (1/18) (\textbf{P12}: \textit{``So \textbf{not trying to get it absolutely right through the LLM}, I think [is] something I found [to be] key in my work with it. \textbf{Let the results be as good as they are and then pick up [...] and keep going on from there}.''}). More strategies are shown in~\autoref{tab:strategies}, along with the challenges or concerns they may help mitigate. These instances showed participants adapting to LLM pitfalls. While many of these strategies could be considered best practices, their necessity puts the onus on users. 
Applications that explicitly support these strategies could help mitigate errors and improve user interactions. We explore this further in the Discussion.

\definecolor{common}{HTML}{E2E2E2}
\definecolor{societal}{HTML}{CCCCCC}
\definecolor{quirks}{HTML}{B8B8B8}
\definecolor{system}{HTML}{A6A6A6}
\definecolor{concerns}{HTML}{909090}
\definecolor{groupline}{gray}{0.4}
\newcommand{\m}{\ensuremath{\vcenter{\hbox{\large$\bullet$}}}}
\newcommand{\e}{\phantom{\m}}

\begin{table*}[t]
\centering
\caption{Strategies for handling LLM responses mapped to applicable challenges. Column codes are expanded below.}
\label{tab:strategies}
\renewcommand{\arraystretch}{1.12}
\setlength{\tabcolsep}{1.5pt}
\scriptsize
\rowcolors{3}{gray!8}{white}
\begin{tabular}{
  !{\color{groupline}\vrule width 1pt}
  >{\raggedright\arraybackslash}m{3.10cm}
  !{\color{groupline}\vrule width 1pt}
  *{4}{>{\centering\arraybackslash}m{0.50cm}|}
  >{\centering\arraybackslash}m{0.50cm}
  !{\color{groupline}\vrule width 1pt}
  *{6}{>{\centering\arraybackslash}m{0.50cm}|}
  >{\centering\arraybackslash}m{0.50cm}
  !{\color{groupline}\vrule width 1pt}
  >{\centering\arraybackslash}m{0.50cm}|
  >{\centering\arraybackslash}m{0.50cm}
  !{\color{groupline}\vrule width 1pt}
  *{2}{>{\centering\arraybackslash}m{0.50cm}|}
  >{\centering\arraybackslash}m{0.50cm}
  !{\color{groupline}\vrule width 1pt}
  *{4}{>{\centering\arraybackslash}m{0.50cm}|}
  >{\centering\arraybackslash}m{0.50cm}
  !{\color{groupline}\vrule width 1pt}
}
\hline
\multicolumn{1}{|c|}{\textbf{Strategy}} &
\multicolumn{5}{c|}{\cellcolor{common}\textbf{Common pitfalls}} &
\multicolumn{7}{c|}{\cellcolor{societal}\textbf{Societal impacts}} &
\multicolumn{2}{c|}{\cellcolor{quirks}\makebox[0pt]{\makecell{\textbf{Model}\\[-1pt]\textbf{quirks}}}} &
\multicolumn{3}{c|}{\cellcolor{system}\makebox[0pt]{\makecell{\textbf{System}\\[-1pt]\textbf{limits}}}} &
\multicolumn{5}{c|}{\cellcolor{concerns}\textbf{User concerns}} \\
\hline
&
\cellcolor{common}\textbf{CP1} & \cellcolor{common}\textbf{CP2} & \cellcolor{common}\textbf{CP3} & \cellcolor{common}\textbf{CP4} & \cellcolor{common}\textbf{CP5} &
\cellcolor{societal}\textbf{SI1} & \cellcolor{societal}\textbf{SI2} & \cellcolor{societal}\textbf{SI3} & \cellcolor{societal}\textbf{SI4} & \cellcolor{societal}\textbf{SI5} & \cellcolor{societal}\textbf{SI6} & \cellcolor{societal}\textbf{SI7} &
\cellcolor{quirks}\textbf{MQ1} & \cellcolor{quirks}\textbf{MQ2} &
\cellcolor{system}\textbf{SL1} & \cellcolor{system}\textbf{SL2} & \cellcolor{system}\textbf{SL3} &
\cellcolor{concerns}\textbf{UC1} & \cellcolor{concerns}\textbf{UC2} & \cellcolor{concerns}\textbf{UC3} & \cellcolor{concerns}\textbf{UC4} & \cellcolor{concerns}\textbf{UC5} \\
\hline
Ask for alternatives & \e & \e & \e & \e & \m & \e & \e & \e & \e & \e & \e & \e & \e & \e & \e & \e & \e & \e & \e & \e & \e & \e \\
\hline
Ask for code instead of chart & \m & \e & \e & \e & \e & \e & \e & \e & \e & \e & \e & \e & \m & \m & \e & \e & \e & \e & \e & \e & \e & \e \\
\hline
Ask for explanations & \e & \e & \e & \e & \e & \m & \e & \e & \e & \m & \m & \m & \e & \e & \e & \e & \e & \e & \e & \e & \e & \e \\
\hline
Ask for pseudocode & \m & \e & \e & \e & \e & \e & \e & \e & \e & \e & \e & \e & \m & \m & \e & \e & \e & \e & \e & \e & \e & \e \\
\hline
Be specific & \m & \e & \m & \e & \e & \e & \e & \e & \e & \e & \e & \m & \e & \e & \m & \e & \e & \e & \e & \e & \e & \e \\
\hline
Break things down & \m & \m & \m & \m & \m & \e & \e & \e & \e & \e & \e & \e & \e & \e & \m & \m & \m & \e & \m & \e & \e & \e \\
\hline
Debug/troubleshoot & \m & \e & \m & \e & \e & \e & \e & \e & \e & \e & \e & \e & \e & \e & \e & \e & \e & \e & \e & \e & \e & \e \\
\hline
Describe data structure & \e & \e & \e & \e & \e & \e & \m & \e & \e & \e & \e & \e & \e & \e & \e & \e & \e & \e & \e & \e & \e & \e \\
\hline
Git repo version control & \e & \m & \e & \e & \e & \e & \e & \e & \e & \e & \e & \e & \e & \e & \e & \e & \e & \e & \e & \e & \e & \e \\
\hline
Give a data sample & \e & \e & \e & \e & \e & \e & \m & \e & \e & \e & \e & \e & \e & \e & \e & \e & \e & \e & \e & \e & \e & \e \\
\hline
Give more guidance & \m & \m & \m & \m & \m & \e & \e & \e & \e & \e & \e & \e & \m & \m & \e & \e & \m & \e & \e & \e & \e & \e \\
\hline
Have a plan & \m & \e & \e & \e & \m & \e & \m & \e & \m & \e & \e & \e & \e & \e & \e & \m & \e & \e & \e & \e & \e & \e \\
\hline
Limit expectations & \e & \m & \e & \m & \e & \e & \e & \e & \e & \e & \e & \e & \e & \e & \e & \e & \e & \e & \e & \e & \e & \e \\
\hline
LLM does the checking & \m & \e & \m & \e & \m & \e & \e & \e & \e & \e & \e & \e & \e & \e & \e & \e & \e & \e & \m & \e & \e & \e \\
\hline
Only use for repetitive tasks & \m & \m & \m & \m & \m & \e & \e & \e & \e & \e & \e & \e & \m & \m & \e & \e & \e & \e & \e & \e & \e & \e \\
\hline
Refine the prompt & \m & \m & \m & \e & \m & \e & \e & \e & \e & \e & \e & \e & \e & \e & \e & \m & \m & \e & \e & \e & \e & \e \\
\hline
Slow down & \m & \e & \e & \e & \e & \e & \e & \e & \e & \e & \e & \e & \e & \e & \e & \e & \e & \e & \e & \e & \e & \e \\
\hline
Start again & \m & \m & \m & \e & \e & \e & \e & \e & \e & \e & \e & \e & \m & \e & \m & \e & \m & \e & \e & \e & \e & \e \\
\hline
Take a break & \m & \e & \e & \e & \e & \e & \e & \e & \e & \e & \e & \e & \e & \e & \e & \e & \e & \e & \e & \e & \e & \e \\
\hline
Use multiple models & \e & \e & \m & \e & \e & \e & \e & \e & \e & \e & \e & \e & \e & \e & \m & \m & \e & \e & \e & \e & \e & \e \\
\hline
\end{tabular}

\vspace{0.45em}
\begin{minipage}[t]{0.32\textwidth}
\scriptsize
\textbf{Common pitfalls}\par
CP1 Errors and hallucinations\par
CP2 Lack of consistency\par
CP3 Misinterpretation\par
CP4 LLM does not know VIS well\par
CP5 Time/effort to correct or finalize

\medskip
\textbf{LLM quirks}\par
MQ1 Rabbit holes\par
MQ2 Verbosity
\end{minipage}\hfill
\begin{minipage}[t]{0.32\textwidth}
\scriptsize
\textbf{Societal impacts}\par
SI1 Bias\par
SI2 Data leak/privacy/cyber\par
SI3 Deepfakes\par
SI4 Environmental impact\par
SI5 Epistemic uncertainty/unknown training data\par
SI6 Lack of trust\par
SI7 Misinformation
\end{minipage}\hfill
\begin{minipage}[t]{0.32\textwidth}
\scriptsize
\textbf{System limits}\par
SL1 Number of tokens\par
SL2 Single point of failure\par
SL3 Some data topics trigger guardrails

\medskip
\textbf{User concerns}\par
UC1 Do not really learn to code\par
UC2 Lack of user skill/training\par
UC3 Limiting/directing thought process\par
UC4 Overreliance\par
UC5 Overtrust
\end{minipage}
\end{table*}

\vspace{1em}
\noindent\textbf{Trust Calibration} (Pipeline Phase: all). Most participants mentioned they do not fully trust the models they use, especially when dealing with sensitive data. Some discussed checking and reviewing LLM outputs. 
Many described a complete lack of trust (\textbf{P11}: \textit{``Yeah, so my base rule is [...]  
\textbf{never trust LLM output}. So [...] I check it with the data.''}
\textbf{P12}: \textit{``But experience tells me in general \textbf{not to trust}, so I always check.''}) or limited trust (\textbf{P1}:\textit{``No, I \textbf{never completely trusted them 100\%}.''}). Frequently, some form of validation was mentioned as a kind of trust lever (\textbf{P16}: \textit{
``\textbf{the only way that I could trust the final results was to review the code}, 
[...] check it line by line to see if the logic is correct.''}). One participant also described a lack of trust in the companies behind LLMs. 
This cautious skepticism and mindful trust manifested in how participants described whether and how they showcase and share LLM outcomes. 4/18 said they only use LLM results internally. Another 4/18 reported openly sharing results, while the majority (10/18) fell in-between; i.e., sharing LLM outputs after modifying or verifying them.
\\\\
\noindent\textbf{Handling Sensitive Data} (Pipeline Phase: Data Collection \& Prep). With sensitive data, 8/18 of participants said they would not use LLMs at all. About 5/18 said they use masking techniques or remove personally identifiable information before using LLMs and 4/18 said they would only share a database schema or description. Other approaches included local or quarantined models (4/18) or using proxy data (3/18). Two participants mentioned relying on LLM privacy settings. 
\\\\
\noindent\textbf{Result Verification} (Pipeline Phase: Review \& Refine). 
Many participants described personal experimentation and strategies to check the model's response. These included looking for visible errors (8/18) (\textbf{P13}: \textit{``what I've done, just in the past anyway with data cleaning, is like the \textbf{sniff test, right? If something doesn't look right}.''}), spot checks (5/18) (\textbf{P11}:\textit{``
\textbf{I take few data points and check}, is it actually valid?''}), and checking descriptive statistics (5/18) (\textbf{P9}: \textit{``I'm going to \textbf{check the min, the max, the middle} and see if that looks like the source data set''}). Other verification techniques included running the LLM-produced code (4/18), looking through the code (2/18) (\textbf{P18}: \textit{``[...] then of course we go with it by \textbf{checking line by line} what it has done.''}), checking against the raw data (3/18), prompting the LLM or a different LLM to check or run tests (3/18), or checking trusted (external) sources (2/18). 
Others reported asking the LLM to generate various types of charts for the same task (2/18), validating via online search (e.g. Google, GitHub)(1/18), and asking the LLM for explanations (1/18).

\subsection{Challenges with LLMs for Data Analytics}
\label{sec:challenges}
While the majority of participant concerns (17/18) revolved around errors (especially hallucinations), several other concerns were raised, from minor quirks to large scale societal impacts. We analyzed and grouped challenges and concerns into five high-level categories. 
\textbf{Common Pitfalls} represent frequent, but often hidden, issues and impacts to the user for given Vis task(s). \textbf{Societal Impacts} represent potential and existing issues which have widespread impact across users and tasks. \textbf{User Concerns} represent broad or long-term concerns about potential negative impacts to the user rather than society at large or a specific task.  \textbf{Model-specific Quirks} are noted model-specific behaviors which may or may not be seen as problems depending on the user and task. Lastly, \textbf{System Limitations} are barriers specific to a given LLM and subscription. In~\autoref{tab:strategies} we crosscut these challenges with applicable participant strategies. 

In the following subsections, we break down where vibe analysis challenges mirror similar concerns in vibe coding and where they diverge. We describe how aspects of data visualization have a unique impact on vibe analysis that differs from vibe coding.

\subsubsection{Challenges prevalent to both vibe analysis and vibe coding}
\label{sec:vibeConcerns}

Most concerns focused on~\textbf{Common Pitfalls}, especially \textbf{hallucinations and errors}. Participants described frequent types of LLM errors, including data ingest, library, filtering, and de-duplication. 11/18 described LLM hallucinations (\textbf{P13}:\textit{ ``And of course that the key part is [accuracy]. [An LLM] can definitely do it faster. \textbf{it hasn't just been accurate. It's hallucinating a lot} in my experiences so far.''}; \textbf{P17}: \textit{``There are times that \textbf{it even generates data points that are outside your data} [...] that is where you have to come in as a human in the loop.''}). Other common pitfalls included a lack of consistency (6/18) (\textbf{P4}: \textit{``Sometimes it gives different answers to the same question and it \textbf{makes it really hard to build reliable workflows because the consistency is not there},''}), lack of Vis-specific guidelines (6/18) (\textbf{P2}: \textit{``I used to make a temperature map; it kind of gave me high temperature in blue and low temperature in red. And \textbf{I think this is pretty, pretty misleading} [...]''}), and the time and effort required to get to a finished product from what the LLM produced (4/18). Armed with this information, UI designers and LLM developers can devise interventions to guide users in each stage of the pipeline. 
\\\\
\textbf{Low Quality} (Pipeline Phase: all). Issues in AI-generated code include subtle bugs, verbosity, security vulnerabilities, and hallucinated APIs~\cite{ge2025survey}. Participants described similar issues specific to Vis, including low chart quality, verbosity, data and specification hallucinations, and data and chart errors during vibe analysis. \textbf{P4:} ``\textit{the most common is when it produces a code that runs but gives some little bit of wrong results like grouping data incorrectly or using a default parameter that I wouldn't have chosen. They usually look very fine at first glance, but then the output doesn't usually like make sense.}'' \textbf{P9}: ``\textit{...there have also been cases where it's like, yeah, Vega-Lite will do this thing. And I'm like Vega-Lite explicitly does not do that thing. And you're basically like arguing with this bot}.'' 
\\\\
\textbf{Technical Debt} (Pipeline Phase: all). Vibe coding can lead to technical debt, unmaintainable code, and buggy or insecure products. It also complicates version control~\cite{pimenova2025good}. Our participants similarly hinted at consistency problems in Vis work, and especially a form of \textit{design debt,} where visual design flaws, inconsistencies, and compromises pile up over time. This impacts analytic reproduceability, comparability, and consistency, since models will generate different responses even in uniform analytic tasks. In some cases this variation may not matter, but many analytic tasks do call for a consistent analytic and design approach, which, even with constraints cannot be guaranteed with LLMs. 
\textbf{P17:} ``\textit{So when that happens, then you have …the first thing you generated is different from the second thing generated. That is some inconsistency.}'' 
\textbf{P18: }``\textit{But the problem that consists of the choice of too many, because then it keeps suggesting you can do this also, you can do that also. And then there are multiple ways of doing the same thing which confuses you.}'' 
For the Vis community, design debt has to be paid eventually through anything from fixing flaws ad hoc to auditing large scale analytics to complete overhauls. With the variety of Vis-specific tools and processes (see Visualization Tasks in Vibe Analytics subsection in Findings) and audience-specific factors (see External Factors subsection in Findings), Vis design debt can be more difficult to unwind and correct than traditional code-based technical debt. Switching between Vis tools like Tableau and Power BI, for example, means updating not just data flows, but Vis components as the availability, compatibility, and translation of design options differ between tools and may be far more rigid than code alone.    
\\\\
\textbf{Interpreting Intent} (Pipeline Phase: all, especially critical to more complex, layered tasks in Planning \& Ideation, Exploration). Vibe coding studies often point to the central impact of ambiguity handling on vibe coding~\cite{ge2025survey}. Ge et al. note that even as 72\% of software defects in production environments originate from misunderstood requirements, state-of-the-art Code LLMs generate code outputs in over 63\% of ambiguous scenarios without seeking clarification~\cite{ge2025survey}. Interpreting and contextualizing analytic intent, central to intelligent analytic systems~\cite{tory2019what}, are analogous challenges in vibe analysis, where the effects of ambiguous requirements (natural language utterances are often underspecified) are compounded by Vis design-specific encoding language. \textbf{P18:} ``\textit{I think the biggest concerns would be explaining the context of [...] why we are making [...] what is it required, and it misinterpreting every simple thing.}''
\\\\
\textbf{Education and Retention Impacts} (Pipeline Phase: Data Collection \& Prep, Exploration). While students learn to verify LLM-generated code, they struggle when asked to integrate their solutions or optimize performance~\cite{geng2025exploring}. 
Some of our participants who were also educators noted that while LLMs give users a shortcut, they do not learn the skills to design good visualizations. \textbf{P14:} ``\textit{...the students could say, OK, I have to [use] chatGPT for that because I don't understand you. I'm not sure it's okay, but it's a little bit sad.}''   
\\\\
\textbf{Productivity Impacts} (Pipeline Phase: Data Collection \& Prep, Exploration). Programmers take less time to understand every line generated by the LLM and more time validating its output against their expectations and mental models~\cite{sarkar2025vibe}. For Vis, participants often note how LLMs speed up content creation, but LLM outputs take time to review and correct. This takes on additional significance when data work is involved, as users have to verify code as in vibe-coding, but also look at data handling and chart design to avoid issues. 
\textbf{P12:} ``\textit{...you really have to pay close attention to to the [model] because at any moment in time, it starts inventing data, it starts adding data, and it sounds really good what it's coming up with, but it's not not factual, it's not the truth.}''~
\\\\
\textbf{Homogeneity} (Pipeline Phase: Planning \& Ideation, Exploration). A common challenge in vibe coding is the limited scope and diversity of the datasets used for training and evaluation~\cite{umama2025llm}. Similarly in vibe analysis, models often revert to the same set of simple charts. In a field known for creativity and innovation, the impact on Vis is to constrain creativity.
\textbf{P3: }``\textit{I found that very limiting in the sense that it almost started to direct my thinking process...}'' 
\\\\
\textbf{Context Window Limitations} (Pipeline Phase: Exploration). 
To deal with context window limitations, vibe coders and vis practitioners sometimes start fresh sessions (see Strategies above).  
Vis practitioners noted the effect of a limited context window on their ability to get meaningful results beyond shorter interactions, which becomes a bigger problem the more complex the analytic requirements. Getting stuck on an increasingly unproductive thread can mean starting from scratch on a multi-layered analytic task, losing hours of work. \textbf{P1} described a routine of getting LLM support for work in Tableau and having to switch from ChatGPT to Copilot and sometimes to Gemini. Each time, work would have to begin again by loading context and data, hoping to be done before the new model's limit is reached. \textbf{P3} described navigating between token limits in ChatGPT and time limits in Claude.
\\\\
\textbf{Model-specific Quirks} (Pipeline Phase: Review \& Refine, Finishing Touches). Verbosity is common to both vibe coding and vibe analysis, though not discussed often in our interviews. This shows up as suboptimal or inefficient code, especially when practitioners use LLMs to set up sites to showcase their work. This added to the difficulty of accurately presenting analytic work and can slow progress or hamper how they share their work, since many Vis practitioners are not web development experts. \textbf{P9:} \textit{``I would end up with these like HTML pages in canvas that were like a few thousand lines long and every change would be like OK, we need to regenerate this whole like text. And that ended up just being slow.''}

\subsubsection{Challenges more prevalent to vibe analysis than vibe coding} 
\label{sec:vibeAnalysisConcerns}

\textbf{Not trained for Vis} (Pipeline Phase: Planning \& Ideation, Exploration, Review \& Refine). Participants noted several issues that expose the lack of Vis-specific training or guidance available to a model. LLMs tested against visualization benchmarks exhibit difficulties with data-dense and deceptive visualizations~\cite{pandey2025benchmarking}, and in practice our participants noted that LLMs regularly fall short on Vis tasks.
 \textbf{P3:} ``\textit{So what Claude was doing is it was generating the grid from left to right even though I was instructing it again and again to, you know, generate the grid from top to bottom.}'' 
\textbf{P7:} ``\textit{But especially for data visualization, I feel that we still haven't reached the ideal, the phase to combine the power of LLM with the needs that we have when we are doing data visualization.}'' 
\textbf{P10: }``\textit{
...table cartograms has actually been a [...] on/off process for me that's actually been going on for six months. It's been more of a how can I actually prompt the LLM to generate this?}'' 
\\\\
\textbf{Hidden code} (Pipeline Phase: Exploration, Review \& Refine). The model's output may not enable immediate or straightforward testing, especially when only the resulting chart is generated. In contrast, vibe coding is generally done either within a development environment where code can be immediately inspected and tested, or users have the option of copying code snippets directly into an interpreter. Vis practitioners must decide how and when to test aspects of the model's code (which is often not automatically provided) along with the output chart(s). \textbf{P4:} ``\textit{I think my biggest concerns are the accuracy and the tiny errors it can make because sometimes it can produce code or analysis that looks pretty much fine, but then it's slightly wrong, and this is like really dangerous [...] when you don't catch it.}'' 
Though hidden code may be a challenge in vibe coding as well, in this case, the programmer is able to see and verify code by default in currently available LLM-based platforms, even when working through a user interface. In contrast, many who use LLMs for chart generation don't inherently see, access, or interact with the code used to generate the chart; for image and chart generation, this code is not typically shared as a default. Secondly, in terms of industry use, software developers using vibe coding usually learn to verify the generated code. Meanwhile, some of our Vis participants do not code, instead working with tools like Excel and Tableau, and so could not verify a model's code.




\subsubsection{Strategy Gaps} As illustrated in Figure~\ref{tab:strategies}, we found that most strategies related to dealing with \textbf{Common Pitfalls}, \textbf{Model-Specific Quirks}, and \textbf{System Limitations}, categories described in the \textit{Challenges with LLMs for Data Analysis} section. The practice of breaking tasks down into steps and providing more context or guidance were among the most cited strategies for these types of issues, along with starting again with a new chat. In contrast, \textbf{Societal Impacts} and \textbf{User Concerns} had few applicable strategies. Some societal impacts may not be completely mitigated -- bias cannot be completely removed, there is no perfect privacy with connected content, and not all forms of uncertainty can be fully captured or calculated. Still, these and other societal concerns could be better acknowledged through clear disclosure by LLM developers, and at the user level when sharing LLM-generated results. This transparency would serve the public interest, users, and could help drive mitigation efforts. User concerns had by far the fewest applicable strategies, representing potential future work. While research to understand and prevent these issues has been conducted, the absence of strategies points to the possibility that developed solutions may be nascent, not well-known, and perhaps not broadly applicable.

\section{Discussion}
\label{sec:discussion}
Here we draw on our interview results to provide recommendations for improving LLM-based Vis. Based on the results, we identify a need for verification support through tool integration, training and resources for practitioners, including best practices and pitfalls to avoid, and design considerations to prevent snowballing design debt. Just as the software development community is grappling with the impact of vibe coding, the Vis community will likewise need to navigate the stochastic nature of LLM-enabled data analysis and visualization.

\subsection{Recommendations for tool developers}
\textbf{Build in verification processes}. Although verification strategies are common among practitioners, they are far from standardized, and different users may have different needs depending on the nature of the data, task, and questions they are tackling. While no single verification technique is universally effective, we found that practitioners rely on combinations of (practical, yet anecdotal) checks and processes to strengthen their analyses and make current LLM tools more robust for data work. 

Almost half of participants noted verification using spot checks or descriptive statistics. Spot checks and multi-model checks provide opportunities to formalize what practitioners are currently doing ad hoc. Incorporating some linting-like processes can provide a more structured, standardized approach to spot checking both data and chart characteristics, while including a comparison across models can help surface both errors and new insights. Spot checking could be automated using more reliable rule-based techniques to complement LLM use. In addition, there is a key research opportunity to develop tools that increase transparency in the actions taken by an LLM on data, to save users the pain of spot checking and digging through code to determine whether and where the LLM made a mistake. Well-designed linting strategies could even nudge users towards noticing that they should verify LLM outputs. 

\noindent\textbf{Incorporate constraints}. P13's example from Finishing Touches represents a good use case for how to provide effective LLM constraints either through prompts or other incorporated guidance. Every minor change to a part of their LLM-generated map should not risk an unwanted change elsewhere.  
Another consideration for incorporating LLMs in Vis tools is the issue of homogeneity, as described in the Challenges section. Developers in particular could consider ways to incorporate, rather than supplant, human creativity. As most of our participants noted their use of LLMs for ideation and visualization design, this represents a major space for innovation. One area for future work could center on how to incentivize and enable human creative input while focusing LLM constraints to mitigate hallucination. On the other hand, if left to models alone, the closed system creates an arbitrary limit on novel design. Domain expertise, design knowledge, and the development and testing of new techniques should be first class citizens in the development process, not secondary to expediency. 

\subsection{Recommendations for the Vis Community}
\textbf{Develop guidance}. LLMs empower users to work beyond their established skillsets. Much like vibe coding, vibe analysis empowers people to quickly iterate through different Vis options, prototype dashboards, and produce other interactive data applications. But we caution that user training is critical. 
The amplifying power brought to bear on everyday Vis tasks could also be used to misinform and manipulate. To mitigate the risks of errors and other LLM pitfalls, we recommend the development of Vis-specific training modules to strengthen AI literacy, as well as verification guidance and processes designed specifically for the use of LLMs for Vis. 
Some examples include guidance about the specific Vis capabilities and limitations of LLMs, best practices and common pitfalls to watch out for, and resources like data and guided problem sets that users can practice with. The strategies described by our expert interviewees can serve as a starting point. 

\noindent\textbf{Provide upskilling resources}. Considering our participants' concerns around job replacement, process changes, and the feeling by some of being compelled to use LLMs, more resources and new approaches are needed to help Vis practitioners navigate the process of incorporating LLMs in their workflows. As some of our interviewees noted, job impacts are perhaps inevitable and likely to come with trade-offs, but whether this includes more job loss or job shift is yet to be determined. 
Opportunities include formalization of training and verification processes, guidance on which aspects of the Vis pipeline are best suited for LLM use, best practices and automated tools for mitigating concerns around data privacy, and approaches to detect and prevent misleading visualization designs and visualization issues that models are known to miss. Training can focus on LLM benefits and how to use them, but should be balanced with a thorough understanding of limitations and risks. The development of practice sets that draw on realistic Vis tasks could be used to help users get comfortable with using LLMs and spotting issues. 


\noindent\textbf{Reckon with design debt}. While technical debt may be important whenever Vis projects rely on LLM-generated code, we identified a separate layer of \textit{design debt} that should be equally concerning. Technical debt encompasses software development decisions that 
may not be immediately harmful, but left to accumulate can render code difficult or impossible to maintain or update~\cite{biazotto2025automating,pimenova2025good}. Design debt are visual design choices (in this case made by an LLM) that may serve an immediate need but can lead to scalability issues, difficulties updating or adding charts or chart features, or higher rates of bugs and errors~\cite{biazotto2025automating}, which in Vis may require more data analysis to identify. Design debt must be dealt with eventually, repaid in patches or overhauls~\cite{biazotto2025automating}, but unlike technical debt, design choices may be less modular than code alone, with license and application-specific limitations or requirements. Ideally, design debt can be prevented before it causes downstream problems by considering broader integration needs. For Vis, this could mean considering both system and audience implications before making choices about format, colors, and other visual designs (marks and channels). Questions about who will maintain and update data, choice of Vis products (e.g., dashboards, interactives), and hosting platforms should be answered early. Limitations of and compatibility between different tools or programming languages used on a project should be considered before switching costs get too high. 

Both P7 and P9 noted use of a Git repository for version control. Though not all Vis practitioners are comfortable using these platforms, some version of a repository specific to Vis projects would help practitioners track changes, revert to previous versions when needed, and allow for traceability and transparency. This type of Vis-focused provenance tracking could also support better sharing and attribution. 

\section{Limitations and Future Work}
Participants represented various backgrounds, sectors, and roles, but there was a skew in country, age, race, and gender. They were not representative of all data workers, but information on how they use LLMs in their work is nonetheless valuable in understanding motivations, challenges, and practices. Participant use of LLMs and their various practices were self-reported and may reflect over- or under-reporting. 
Future research can extend this work to a broader sample, including data workers from other professions who may have lower visualization expertise and/or see data analytics as a secondary part of their role. 

Our study focused on LLMs due to their broad use, availability, and ability to handle a broad range of Vis tasks. 
Future work can also re-assess to see how findings evolve as the technology advances.
Other opportunities for future work include identifying best-fit and user preferences for LLMs and traditional Vis tools, to guide how to split Vis functions in a hybrid process, and 
investigating design debt in Vis and how to best prevent or mitigate it. 

\section{Conclusion}
We conclude with a call to the visualization community to think critically about where and how we apply LLMs and how we learn and train others to use them to present data. LLMs offer a tremendous opportunity to enable users to quickly and easily explore and visualize data. 
But this power has so far been applied without the requisite responsibility, and with inadequate guardrails. The next big question for the community may not be \textit{whether} to use LLMs for Vis, as this may soon become a moot point, but instead how to effectively prepare users (from lay users to advanced practitioners) to use them effectively. 

In short, LLMs may speed up Vis work, but they may also increase verification burden, weaken skill formation, narrow design exploration, and create downstream maintenance and quality problems. For a community that values clear and accurate data representation, this presents a challenge, but also opportunities to innovate and adapt.

We close with a quote about the importance of goal-driven exploration and experimentation, recognizing that while there are some aspects of data analysis and visualization that greatly benefit from the use of LLMs, there may be others that should not be offloaded:
    \textit{``Because learning does not consist only of knowing what we must or we can do but also of knowing what we could do and perhaps should not do.''} - Umberto Eco


\section{ACKNOWLEDGMENT}

We are grateful to Jane Adams and Racquel Fygenson for their input. 
This material is based upon work supported by Northeastern University's Khoury College of Computer Sciences and the National Science Foundation (NSF) CISE Graduate Fellowships under Grant No. 2313998. Any opinions, findings, and conclusions or recommendations expressed in this material are those of the author(s) and do not necessarily reflect the views of the NSF.

\bibliographystyle{ieeetr}
\bibliography{References}

@misc{qian2024evolution,
	title = {The Evolution of {LLM} Adoption in Industry Data Curation Practices},
	url = {http://arxiv.org/abs/2412.16089},
	doi = {10.48550/arXiv.2412.16089},
	number = {{arXiv}:2412.16089},
	publisher = {{arXiv}},
	author = {Qian, Crystal and Liu, Michael Xieyang and Reif, Emily and Simon, Grady and Hussein, Nada and Clement, Nathan and Wexler, James and Cai, Carrie J. and Terry, Michael and Kahng, Minsuk},
	urldate = {2025-01-10},
	year = {2024},
	eprinttype = {arxiv},
	eprint = {2412.16089 [cs]},
}

@inproceedings{tory2019what,
	address = {Vancouver, {BC}, Canada},
	title = {Do What {I} Mean, Not What {I} Say! Design Considerations for Supporting Intent and Context in Analytical Conversation},
	rights = {https://ieeexplore.ieee.org/Xplorehelp/downloads/license-information/{IEEE}.html},
	isbn = {978-1-7281-2284-7},
	url = {https://ieeexplore.ieee.org/document/8986918/},
	doi = {10.1109/VAST47406.2019.8986918},
	eventtitle = {2019 {IEEE} Conf. Visual Analytics Science and Technology ({VAST})},
	pages = {93--103},
	booktitle = {2019 {IEEE} Conf. Visual Analytics Science and Technology ({VAST})},
	publisher = {{IEEE}},
	author = {Tory, Melanie and Setlur, Vidya},
	urldate = {2025-06-03},
	year = {2019},
}

@misc{pandey2025benchmarking,
	title = {Benchmarking Visual Language Models on Standardized Visualization Literacy Tests},
	url = {http://arxiv.org/abs/2503.16632},
	doi = {10.48550/arXiv.2503.16632},
	number = {{arXiv}:2503.16632},
	publisher = {{arXiv}},
	author = {Pandey, Saugat and Ottley, Alvitta},
	urldate = {2025-03-28},
	year = {2025},
	eprinttype = {arxiv},
	eprint = {2503.16632 [cs]},
}

@misc{yao2024llm,
      title={{LLM} Lies: Hallucinations are not Bugs, but Features as Adversarial Examples}, 
      author={Jia-Yu Yao and Kun-Peng Ning and Zhen-Hui Liu and Mu-Nan Ning and Yu-Yang Liu and Li Yuan},
      year={2024},
      eprint={2310.01469},
      archivePrefix={arXiv},
      primaryClass={cs.CL},
      url={https://arxiv.org/abs/2310.01469}, 
      doi={10.48550/arXiv.2310.01469
}
}

@ARTICLE{crisan2021passing,
  author={Crisan, Anamaria and Fiore-Gartland, Brittany and Tory, Melanie},
  journal={IEEE Trans. Visualization and Computer Graphics}, 
  title={Passing the Data Baton: A Retrospective Analysis on Data Science Work and Workers}, 
  year={2021},
  volume={27},
  number={2},
  pages={1860-1870},
  doi={10.1109/TVCG.2020.3030340}}

@article{schetinger2023doom,
author = {Schetinger, V. and Di Bartolomeo, S. and El-Assady, M. and McNutt, A. and Miller, M. and Passos, J. P. A. and Adams, J. L.},
title = {Doom or Deliciousness: Challenges and Opportunities for Visualization in the Age of Generative Models},
journal = {Computer Graphics Forum},
volume = {42},
number = {3},
pages = {423-435},
doi = {https://doi.org/10.1111/cgf.14841},
url = {https://onlinelibrary.wiley.com/doi/abs/10.1111/cgf.14841},
eprint = {https://onlinelibrary.wiley.com/doi/pdf/10.1111/cgf.14841},
year = {2023}
}

@article{kosminsky2019belief,
   author = "Kosminsky, Doris and Walny, Jagoda and Vermeulen, Jo and Knudsen, Søren and Willett, Wesley and Carpendale, Sheelagh",
   title = "Belief at first sight", 
   journal= "Information Design Journal",
   year = "2019",
   volume = "25",
   number = "1",
   pages = "43-55",
   doi = "https://doi.org/10.1075/idj.25.1.04kos",
   url = "https://www.jbe-platform.com/content/journals/10.1075/idj.25.1.04kos",
   publisher = "John Benjamins",
   issn = "0142-5471",
   type = "Journal Article",
  }

@misc{bindley2026tech,
author = {Bindley, Katherine and Blunt, Katherine},
title = {Tech Firms Aren't Just Encouraging Their Workers to Use AI. They're Enforcing It.},
publisher = {Wall Street Journal},
year = {2026},
date = {25 Feb. 2026},
url = {https://www.wsj.com/tech/ai/tech-firms-arent-just-encouraging-their-workers-to-use-ai-theyre-enforcing-it-d43ebf84?mod=tech_trendingnow_article_pos2},
}

@misc{sarkar2025vibe,
      title={Vibe coding: programming through conversation with artificial intelligence}, 
      author={Advait Sarkar and Ian Drosos},
      year={2025},
      eprint={2506.23253},
      archivePrefix={arXiv},
      primaryClass={cs.HC},
      url={https://arxiv.org/abs/2506.23253}, 
}

@misc{geng2025exploring,
      title={Exploring Student-{AI} Interactions in Vibe Coding}, 
      author={Francis Geng and Anshul Shah and Haolin Li and Nawab Mulla and Steven Swanson and Gerald Soosai Raj and Daniel Zingaro and Leo Porter},
      year={2025},
      eprint={2507.22614},
      archivePrefix={arXiv},
      primaryClass={cs.HC},
      url={https://arxiv.org/abs/2507.22614}, 
}

@misc{ge2025survey,
      title={A Survey of Vibe Coding with Large Language Models}, 
      author={Yuyao Ge and Lingrui Mei and Zenghao Duan and Tianhao Li and Yujia Zheng and Yiwei Wang and Lexin Wang and Jiayu Yao and Tianyu Liu and Yujun Cai and Baolong Bi and Fangda Guo and Jiafeng Guo and Shenghua Liu and Xueqi Cheng},
      year={2025},
      eprint={2510.12399},
      archivePrefix={arXiv},
      primaryClass={cs.AI},
      url={https://arxiv.org/abs/2510.12399}, 
}

@misc{pimenova2025good,
      title={Good Vibrations? A Qualitative Study of Co-Creation, Communication, Flow, and Trust in Vibe Coding}, 
      author={Veronica Pimenova and Sarah Fakhoury and Christian Bird and Margaret-Anne Storey and Madeline Endres},
      year={2025},
      eprint={2509.12491},
      archivePrefix={arXiv},
      primaryClass={cs.SE},
      url={https://arxiv.org/abs/2509.12491}, 
}

@misc{chou2025building,
      title={Building Software by Rolling the Dice: A Qualitative Study of Vibe Coding}, 
      author={Yi-Hung Chou and Boyuan Jiang and Yi Wen Chen and Mingyue Weng and Victoria Jackson and Thomas Zimmermann and James A. Jones},
      year={2025},
      eprint={2512.22418},
      archivePrefix={arXiv},
      primaryClass={cs.SE},
      url={https://arxiv.org/abs/2512.22418}, 
}

@ARTICLE{umama2025llm,
  author={Umama and Usman Danyaro, Kamaluddeen and Nasser, Maged and Zakari, Abubakar and Abdullahi, Shamsu and Khanzada, Atika and Muntasir Yakubu, Muhammad and Shoaib, Sara},
  journal={IEEE Access}, 
  title={LLM-Based Code Generation: A Systematic Literature Review With Technical and Demographic Insights}, 
  year={2025},
  volume={13},
  number={},
  pages={194915-194939},
  doi={10.1109/ACCESS.2025.3631952}}

@misc{nolan2026top,
title ={Top engineers at Anthropic, OpenAI say {AI} now writes 100\% of their code—with big implications for the future of software development jobs},
author = {Nolan, Beatrice},
year = {2026},
url ={https://fortune.com/2026/01/29/100-percent-of-code-at-anthropic-and-openai-is-now-ai-written-boris-cherny-roon/}}

@book{riche2018analysis,
    author = {Riche, Nathalie Henry and Hurter, Christophe and Diakopoulos, Nicholas and Carpendale, Sheelagh },
    title = {Data-Driven Storytelling},
    chapter = {From Analysis to Communication},
    publisher = {A K Peters/CRC Press},
    year = {2018}
}

@inproceedings{mcnutt2020surfacing,
author = {McNutt, Andrew and Kindlmann, Gordon and Correll, Michael},
title = {Surfacing Visualization Mirages},
year = {2020},
isbn = {9781450367080},
publisher = {Association for Computing Machinery},
address = {New York, NY, USA},
url = {https://doi.org/10.1145/3313831.3376420},
doi = {10.1145/3313831.3376420},
booktitle = {Proc. 2020 CHI Conf. Human Factors in Computing Systems},
pages = {1–16},
numpages = {16},
location = {Honolulu, HI, USA},
series = {CHI '20}
}

@misc{biazotto2025automating,
      title={Automating Technical Debt Management: Insights from Practitioner Discussions in Stack Exchange}, 
      author={João Paulo Biazotto and Daniel Feitosa and Paris Avgeriou and Elisa Yumi Nakagawa},
      year={2025},
      eprint={2502.03153},
      archivePrefix={arXiv},
      primaryClass={cs.SE},
      url={https://arxiv.org/abs/2502.03153}, 
}

@misc{shaw2026thinking,
author = {Shaw, Steven D and Nave, Gideon},
title ={Thinking—Fast, Slow, and Artificial: How {AI} is Reshaping Human Reasoning and the Rise of Cognitive Surrender},
year = {2026},
date = {January 11, 2026},
url = {https://doi.org/10.31234/osf.io/yk25n_v1},
doi = {10.31234/osf.io/yk25n_v1}}

@article{naeem2023step,
title = "A Step-by-Step Process of Thematic Analysis to Develop a Conceptual Model in Qualitative Research",
author = "Muhammad Naeem and Wilson Ozuem and Kerry Howell and Silvia Ranfagni",
year = "2023",
month = nov,
day = "8",
doi = "10.1177/16094069231205789",
language = "English",
volume = "22",
pages = "1--18",
journal = "International Journal of Qualitative Methods",
issn = "1609-4069",
publisher = "SAGE",

}

\begin{IEEEbiography}{Shani C Spivak}{\,} is a PhD candidate at Northeastern University and the primary contact of this paper (spivak.s@northeastern.edu). Her research lies at the intersection of Vis, AI ethics and governance, AI literacy and accessibility, and human-computer interaction.
\end{IEEEbiography}

\begin{IEEEbiography}{Aditi Krishna}{\,} is a PhD student at Northeastern University, Boston, Massachusetts, USA. Her research interests include human-AI interaction and responsible AI.  Contact her at krishna.ad@northeastern.edu.
\end{IEEEbiography}

\begin{IEEEbiography}{Mahsan Nourani} {\,} is an Assistant Research Professor with Northeastern University, Portland, Maine, USA. Her research interests include human-centered AI and visual analytics. Nourani received her PhD in Computer Science from University of Florida. Contact her at m.nourani@northeastern.edu.
\end{IEEEbiography}

\begin{IEEEbiography}{Melanie Tory} {\,} is Professor of the Practice with Northeastern University, Oakland, California, USA. Her research interests include visualization and human-AI interaction. Tory received her PhD in Computer Science from Simon Fraser University. Contact her at m.tory@northeastern.edu.
\end{IEEEbiography}

\end{document}


\renewcommand\thefigure{\thesection.\arabic{figure}}  
\maketitle
\setcounter{figure}{0} 
\setcounter{table}{0}
\renewcommand{\thetable}{A.\arabic{table}}

\section{Participant Recruitment}
DVS is a cross-functional, tool-agnostic community for sharing ideas, learning, and exchanging best practices for data visualization design. This community operates a research recruitment program where members can opt into a dedicated list of opportunities to participate as research subjects in academic studies, enabling researchers to easily connect with data visualization practitioners. See~\autoref{tab:demographics} for more details about participants.

\definecolor{Iron}{rgb}{0.85,0.862,0.862}
\begin{table*}h
\centering
\caption{Information about participant demographics and roles. (NP = Non-profit; Black/AA = Black or African American). These categories did not inform analysis, and are included only to describe sample composition.}
\label{tab:demographics}
\begin{tblr}{
  width = \linewidth,
  colspec = {Q[75,l]Q[85,l]Q[110,l]Q[133,l]Q[130,l]Q[200,l]Q[150,l]},
  row{even} = {Iron},
  cell{1}{1} = {font=\bfseries},
  cell{1}{2} = {font=\bfseries},
  cell{1}{4} = {font=\bfseries},
  cell{1}{5} = {font=\bfseries},
  cell{1}{6} = {font=\bfseries},
  cell{1}{7} = {font=\bfseries},
  hline{2} = {-}{0.08em},
  hline{20} = {-}{0.08em},
}
\textbf{P\#} & Age & \textbf{Gender}   & Race                      & Country & Sector                                                           & Org. Role                                \\
P1          & 51-60     & Female            & Asian                     & USA     & Private; NP                                       & Analyst                                 \\
P2          & 21-30     & Male              & Asian                     & USA     & Academia                                                         & Analyst                                 \\
P3          & 31-40     & Male              & Asian                     & Canada  & Government; \newline Private; NP           & Designer                                 \\
P4          & 21-30     & Male              & Black/AA & USA     & Private; Academia                                         & Analyst                                 \\
P5          & 31-40     & Female            & Black/AA & USA     & NP; Academia                                             & Analyst                                  \\
P6          & 51-60     & Male              & White                     & Denmark & Government; \newline Private;           & Developer / \newline Scientist       \\
P7          & 31-40     & Male              & Hispanic                  & Brazil  & NP, Academia; \newline Multilateral orgs         & Journalist                            \\
P8          & 21-30     & Female            & Asian                     & India   & Private                                     & Developer                             \\
P9          & 31-40     & Male              & White                     & USA     & Government; \newline Private; NP           & Designer                                \\
P10         & 51-60     & Male              & White                     & Canada  & Government; \newline Private; Academia             & Leadership \\
P11         & 31-40     & Male              & Asian                     & USA     & Government                                       & Scientist                               \\
P12         & 41-50     & Female            & White                     & Austria & NP                                                       & Analysis, \newline Visualization          \\
P13         & 51-60     & Prefer not \newline to say & White                     & USA     & NP                                                       & Evaluator                         \\
P14         & 31-40     & Male              & Hispanic                  & Mexico  & Government; Private; \newline NP; Academia & Analyst                                  \\
P15         & 31-40     & Male              & White                     & USA     & Private                                                 & Designer                                \\
P16         & 31-40     & Male              & White                     & USA     & Academia                                                         & Scientist                               \\
P17         & 31-40     & Male              & Black/AA & Nigeria & Private; NP                                       & Analyst                                  \\
P18         & 31-40     & Male              & Asian                     & India   & Academia                                                         & Engineer                               
\end{tblr}
\end{table*}

\section{Survey Development}
We developed survey questions in three groups: 1) questions about participants' data visualization experience and expertise, 2) questions about their use of LLMs, and 3) demographic questions.
The visualization expertise questions were adapted from the Data Visualization Society (DVS) annual survey to assess participants' roles, sectors, audiences, preferred tools, and confidence in their abilities to create effective visualizations. Questions about LLM usage explored participants' frequency of use, integration points within their workflows, specific examples of LLM use, verification practices, and confidence in their abilities to get meaningful results from LLMs. We include more general questions along with Vis-specific ones because they show comfort with/familiarity with LLMs in general as context for their use in their vis work. Finally, demographic questions followed the format used in the annual DVS survey and were included at the end to minimize response bias. 

The survey was shared through DVS and deployed via Qualtrics. Participants were shown an information page first to provide them with general information about the study and confirming their privacy rights and consent.

\subsection{Questions about Vis Expertise}
\hfill \\
\noindent 1) As related to data visualization, which of the following describe your role in the past year? (multi-select)
\begin{itemize}
\item Freelance/Consultant/Independent contractor
\item Position in an organization
\item Non-compensated data visualization hobbyist
\item Student in a degree program at a college or university
\item Academic/Teacher
\item Passive income from data visualization related products
\item Prefer not to answer
\item Other (fill in)
\end{itemize}

\noindent 2) How confident are you in your ability to create effective data visualizations that accurately represent complex information?
\begin{itemize}
\item 1 - Not at all confident
\item 2 - 
\item 3 - Slightly confident
\item 4 - 
\item 5 - Moderately confident
\item 6 - 
\item 7 - Very confident
\item 8 - 
\item 9 - Extremely confident
\end{itemize}

\noindent 3) In which of the following sectors does your organization operate? (multi-select)
\begin{itemize}
\item Public sector (government)
\item Private sector
\item Non-profit
\item Academia
\item Other (fill in)
\item Not Applicable
\end{itemize}

\noindent 4) What best describes your role in your organization? (Select One)
\begin{itemize} 
\item Analyst
\item Cartographer
\item Designer
\item Developer
\item Engineer
\item Journalist
\item Leadership (Manager, Director, VP, etc.)
\item Scientist
\item Teacher
\item Other (fill in)
\item Not Applicable
\end{itemize}

\noindent 5) Who do you make data visualizations for? (multi-select)
\begin{itemize}
\item General public
\item Engineers 
\item Medical professionals
\item Policymakers
\item Designers
\item Students
\item Administrative personnel
\item Yourself
\item Analysts
\item Scientists
\item Product or project managers
\item Executives
\item Researchers
\item Educators
\item Other data visualization enthusiasts
\item Other (fill in)
\end{itemize}

\noindent 6) Which tools do you tend to use for data visualization? (multi-select)
\begin{itemize}
\item ArcGIS
\item Canva
\item Canvas
\item D3
\item Datawrapper
\item Excel
\item Figma
\item Flourish
\item ggplot2
\item Gephi
\item GoogleDataStudio
\item GoogleSheets
\item Highcharts
\item Illustrator
\item Leaflet
\item Looker
\item Mapbox
\item Matplotlib
\item Observable
\item P5orProcessing
\item PenPaper
\item PhysicalMaterials
\item Plotly
\item PowerBI
\item PowerPoint
\item Python
\item QGIS
\item Qlik
\item R
\item RAWGraphs
\item React
\item Svelte
\item Tableau
\item Vega
\item Vue
\item WebComponents
\item WebGL
\item Other
\end{itemize} 

\subsection{Questions about LLM use}
\hfill \\
\noindent 7) For what part of the data visualization process do you use Large Language Models (LLMs)? (multi-select)
\begin{itemize}
\item Ideation
\item Domain understanding/exploration
\item Data wrangling
\item Data familiarization / exploratory data analysis
\item Inspiration for visualization designs
\item Finish/polish
\item Iconography/image generation
\item Other (please specify)
\item I don't use LLMs for this
\end{itemize}

\noindent 8) Please provide an example (open text)

\noindent 9) About how often would you say you use LLMs (ChatGPT, Gemini, Claude, LLama, etc.) in general (work and non-work)?
\begin{itemize}
\item I use them every day
\item I use them more than once a week
\item I use them monthly
\item I use them quarterly
\item I've used them only a few times
\item Never
\end{itemize}

\noindent 10) Do you verify the response when using LLMs?
\begin{itemize}
\item Yes
\item No
\item Sometimes
\end{itemize}

\noindent 10a) (if yes or sometimes) How do you verify the response? (check all that apply)
\begin{itemize}
\item Spot check some values or fields
\item Check the result against a reference
\item Work through the solution manually
\item Other
\end{itemize}

\noindent 10b)(if yes or sometimes) Please provide an example of how you verified an LLM response
(open text)\hfill \\

\noindent 11) How confident are you in your ability to get meaningful results from an LLM?
\begin{itemize}
\item 1 - Not at all confident
\item 2 - 
\item 3 - Slightly confident
\item 4 - 
\item 5 - Moderately confident
\item 6 - 
\item 7 - Very confident
\item 8 - 
\item 9 - Extremely confident
\end{itemize}

\noindent 12) Which LLMs have you used? (check all that apply)
\begin{itemize}
\item BERT (Google)
\item BLOOM (HuggingFace)
\item ChatGPT (OpenAI)
\item Claude (Anthropic)
\item Gemini (Google)
\item Llama (Meta)
\item Mixtral (Mistral AI)
\item PanGu- (Huawei)
\item Other (write in)
\end{itemize}
\noindent 13) What do you find most helpful in learning to use LLMs? (check all that apply)
\begin{itemize}
\item Books
\item Classes 
\item Workshops
\item Examples
\item Research papers
\item Video tutorials
\item Working with practitioners
\item mentoring/teaching others
\item Podcasts
\item Other (write in)
\item I don't use any learning materials for this
\end{itemize}

\noindent 14) Based on the previous questions in this survey, what is the main topic we're studying?
\begin{itemize}
\item How people use LLMs for writing and editing documents
\item How people use LLMs for creating charts and graphs
\item How people use LLMs for coding and software development
\item How people use LLMs for language translation tasks
\item Other (write-in)
\end{itemize}

\noindent 15) If you are willing to participate in a longer (about one hour), paid interview session about your use of LLMs in data visualization development, please provide your full name and email here. Names and emails will be removed prior to publication to maintain participant anonymity. (open text)

\subsection{Demographic Questions}  \hfill \\
\noindent 16) What country do you live in?
(open text)

\noindent 17) What is your gender?
\begin{itemize}
\item Female
\item Male
\item Nonbinary
\item Prefer not to say
\item Prefer to self describe (fill in)
\end{itemize}

\noindent 17) What is your age range?
\begin{itemize}
\item <21
\item 21-30
\item 31-40
\item 41-50
\item 51-60
\item 61-70
\item 71-80
\item 81-90
\item >90
\end{itemize}
\noindent 18) Race/Ethnicity (check all that apply)
\begin{itemize}
\item American Indian or Alaska Native (A person having origins in any of the original peoples of North and South America (including Central America) and who maintains tribal affiliation or community attachment.)
\item Asian (A person having origins in any of the original peoples of the Far East, Southeast Asia, or the Indian subcontinent including, for example, Cambodia, China, India, Japan, Korea, Malaysia, Pakistan, the Philippine Islands, Thailand, and Vietnam.)
\item Black or African American (A person having origins in any of the Black racial groups of Africa.)
\item Hispanic (A person of Cuban, Mexican, Puerto Rican, South or Central American, or other Spanish culture or origin regardless of race.)
\item Native Hawaiian or Other Pacific Islander (A person having origins in any of the original peoples of Hawaii, Guam, Samoa, or other Pacific Islands)
\item White (A person having origins in any of the original peoples of Europe, the Middle East, or North Africa.)
\end{itemize}

\subsection{Interview Guide} 
[Interviewee greets Participant]

Provide interviewee with Interview Information Sheet and give them time to review.

“Thank you for agreeing to participate in this interview. We are conducting this study to better understand how data visualization practitioners use Large Language Models, such as ChatGPT or Gemini. We are most interested in hearing about your experiences using LLMs for data work, especially for data visualization creation, but this may also include data prep, exploration, etc. This study has been approved by an IRB. After reviewing the information sheet, do you have any questions about the study or IRB?” 

“As a reminder, your participation is voluntary, and you may withdraw at any time.” 

“This interview will be recorded for research purposes, and your confidentiality will be maintained. We will use an ID number to identify information related to this interview and anonymize the content. We will transcribe the interview and then de-identify the transcripts. The recordings will be removed at the conclusion of the project. May I have your consent to record? You may turn off your video if you like.”

[Interviews were semi-structured and roughly followed the question order below.]


1) Briefly tell me about your job role and how you use data in your work.

2) In which parts of the data visualization pipeline have you used LLMs (ideation, data cleaning / prep, statistics, exploratory analysis, visualization creation, etc.)?

3) How did you first start using LLMs for data visualization? Was there a specific goal or use-case you had in mind (e.g. for data exploration)?

4) Considering LLM-generated data outputs like brainstorming ideas, data processing code, or visualizations, do you often share with others, or is it for personal use?

5) Which LLM-based tools do you prefer for visualization tasks, and why? Do you think any are better than the others?

6) How do you decide when to use LLMs versus other visualization tools?

7) Do you have go-to prompts or strategies that work well for data or visualization tasks with LLMs?

8) How has incorporating LLMs changed your traditional visualization workflow or process? (Possible follow-up: can you give me some examples?)

9) Discussing an example:
Can you give us an example of how you use LLMs in your data work?

Please show us a past transcript of the process if you can. Feel free to show us a demo dataset if you can't share the data with us.

Please walk us through your process step by step.

Was there a time when you were able to discover or generate specific insights using LLMs that you might not have been able to otherwise?

10) Challenges with LLMs: What do you do when the LLM doesn't understand your visualization or data prompt or misinterprets it? Please provide an example.

11) Do you usually have to go through iterations with the LLM before getting a satisfactory result for visualizations or data work? If yes, ask for an example.

12) How do you determine when to trust an LLM-generated visualization or other data product versus when to be more skeptical? (Possible follow up: What additional information, if any, would you seek to determine the trustworthiness of an LLM response?)

13) Have you encountered situations where the LLM produced misleading or incorrect visualizations or data products? (If yes, ask for an example.) How did you handle that?

14) How do you handle sensitive or proprietary data when using LLMs for data work?

15) What are your biggest concerns or limitations when using LLMs for data work?

16) Do you have any concerns about being replaced by LLMs?

17) Would you like to share any final thoughts or comments that you didn't get a chance to discuss? 

Thank you very much for your participation. 
Your gift card will be emailed to you at the email address you provided.

\section{Visualization Tools Used}
\begin{figure}
    \centering
    \includegraphics[width=1\linewidth]{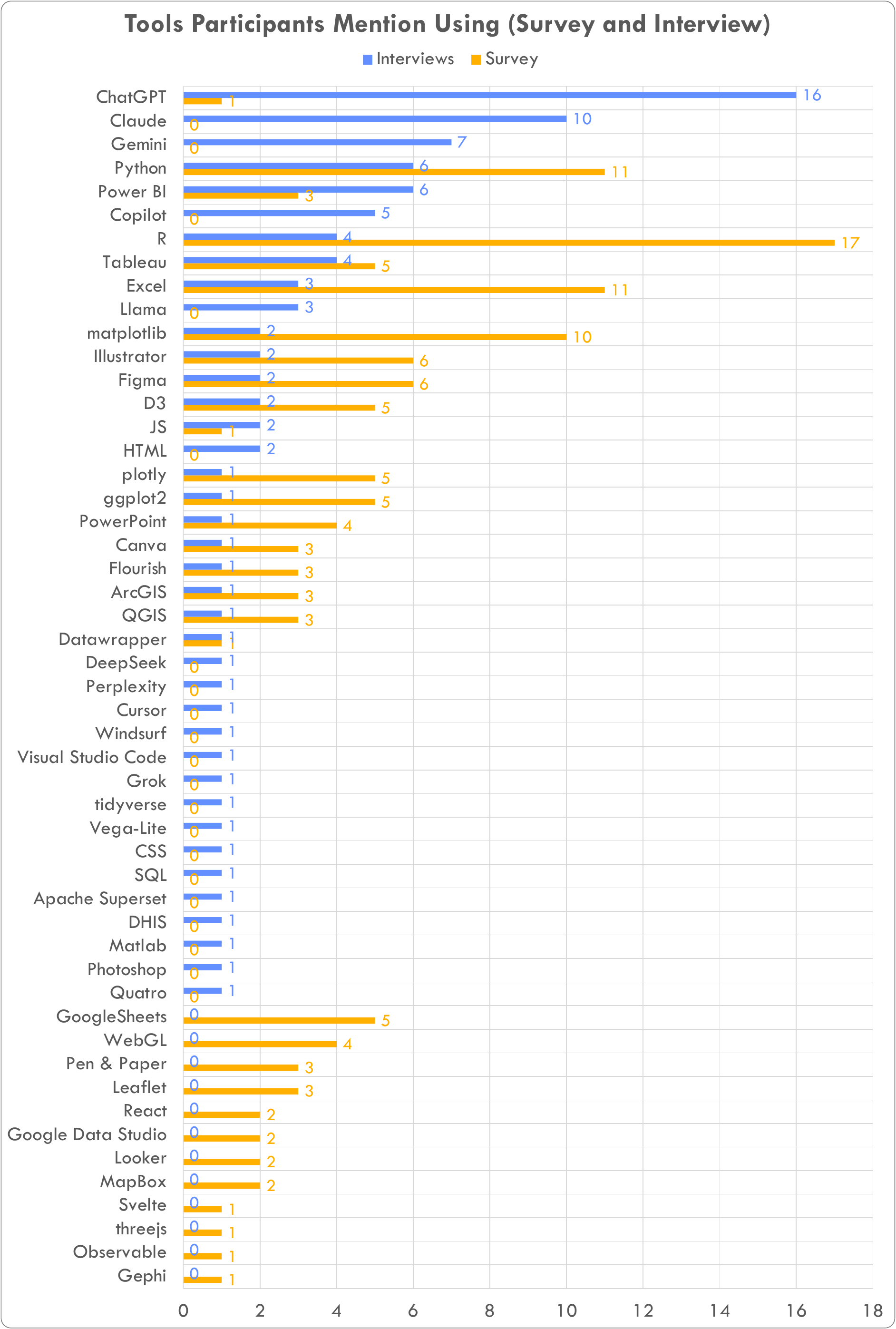}
    \caption{Count of the tools participants reported using vis the survey (in yellow) and via interviews (in blue).}
    \label{fig:tools}
\end{figure}

\textbf{Figure \ref{fig:tools}} shows the tools participants reported using, color coded by whether they discussed use of these tools via the intake survey or throughout the interview.

